\documentclass[preprint,tightenlines,preprintnumbers,showpacs,nofootinbib]{revtex4-1}
\usepackage{epsf}

\newcommand \MSbar {$\overline{{\rm MS}}$}

\newcommand \polepart {\loarrow{\cal P}}

\newcommand \tr {\mathop{\rm tr}}

\newbox \JCCHoldBox
\newdimen \JCCLower
\newcommand \epscenterbox [2]%
{    
     \epsfxsize=#1%
     \setbox \JCCHoldBox = \hbox{\epsfbox{#2}}%
     \JCCLower = 0.5\ht\JCCHoldBox
     \advance \JCCLower by -0.3ex%
     \lower \JCCLower \box\JCCHoldBox
}

\begin{document}
\draft

\preprint{
        \parbox{1.5in}{%
           \noindent
           hep-ph/9806259 \\
           PSU-TH/198
        }
}

\title{Hard-scattering factorization with heavy quarks: A general
       treatment}

\author{J.C. Collins}
\address{
         Penn State University,
         104 Davey Lab, University Park PA 16802, U.S.A.
        }

\date{23 January 2015}

\begin{abstract}%
    A detailed proof of hard-scattering factorization is given with the
    inclusion of heavy quark masses. Although the proof is
    explicitly given for deep-inelastic scattering, the methods
    apply more generally.  The power-suppressed corrections to the
    factorization formula are uniformly suppressed by a power of
    $\Lambda /Q$, independently of the size of heavy quark masses, $M$,
    relative to $Q$.
\end{abstract}


\maketitle

\section{Introduction}
\label{sec:introduction}

A correct treatment of heavy quarks in higher-order perturbative QCD
calculations is important
\cite{F2c,photo,Gluck.Reya,DEN,had.heavy,ACOT,other,BMSvN,MRRS,RT2,RT1}
to precision phenomenology.  Among the reasons is the fact that a
substantial fraction of the deep-inelastic cross section at HERA is in
heavy quark production.  Moreover, this occurs in a region where the
heavy quark masses are not necessarily negligible with respect to the
large momentum scales in the problem (like $Q$).

However, there is a considerable confusion
\cite{Gluck.Reya,ACOT,other,BMSvN,MRRS,RT2,RT1} about what constitute correct
methods for treating heavy quarks.  Some of the difficulties occur
because many treatments
assume that quarks either are so light that their masses are
negligible with respect to $Q$ or have masses that are of order $Q$,
where $Q$ denotes a typical scale for the hard-scattering process
under discussion.  One has to be able to handle the intermediate
region, where $Q$ is somewhat larger than a quark mass but not
enormously much larger.

Even when $Q$ is much larger than all quark masses, the
intermediate region must still be treated, because evolution
equations are used to obtain the strong coupling, the parton
densities, and the fragmentation functions from starting values
specified at scales of a few GeV.
The symptoms of this issue are
the different and apparently incompatible ``matching conditions''
that have been proposed.\footnote{
        See for example \cite{MRRS,ACOT}.
}

In this paper I will give a relatively
simple and general proof of factorization
including the effects of heavy quarks.  The only issue that will not
be treated is the cancellation of soft gluons, an issue which is
essentially orthogonal to the ones which are causing problems.  The
key ingredient is the observation that the short-distance coefficient
functions (``Wilson coefficients'') can legitimately be calculated with
the quark masses left non-zero.  Previous work with Aivazis, Olness
and Tung \cite{ACOT} and others \cite{other} has used this method;
what is new is the complete and detailed all-orders proof.

This first main characteristic of the method, that quark masses are
retained when necessary in the calculations of the coefficient
functions, enables factorization to be valid when the masses of quarks
are non-negligible with respect to the large scale $Q$ of the hard
scattering.  Hence the method avoids the normal problem when the
\MSbar{} scheme is used with massless Wilson coefficients, that there
are uncontrolled corrections of order a power of $M/Q$, where $M$ is a
heavy quark mass.

The second main characteristic is that the renormalization and
factorization scheme consists of a series of subschemes labeled by
the number of ``active quark flavors'', $n_{A}$.  This is simply a
generalization of the Collins, Wilczek and Zee (CWZ)
scheme \cite{CWZ} that is in standard use
\cite{PDG} for the QCD coupling $\alpha _{s}$.  When discussing the
numerical values of parton densities, it is necessary to specify the
number of active flavors that is used in their definition, just as in
the case of the coupling.

The subschemes with different numbers of active flavors are useful in
different ranges of physical scales, but with overlapping ranges of
validity.  Since the subschemes are related by definite matching
conditions \cite{coupling-match,pdf-match}, the choice of the number
of active flavors does not result in any more indefiniteness in the
physical predictions than does the freedom to choose a scheme or a
value of the renormalization/factorization scale.

At first sight, the use of a sequence of subschemes instead of a
single scheme appears rather baroque.  However, it is in fact the
simplest implementation of mass-dependent factorization
\cite{perfect}.  We require that the schemes implement decoupling
\cite{AC} of heavy quarks when appropriate, and that they implement
the closest possible scheme to the mass-independent \MSbar{} scheme,
which is commonly used for most perturbative QCD calculations.  If one
did not have a sequence of schemes, it would be necessary to have
mass-dependent evolution equations.  The CWZ scheme does have
mass-dependent evolution in the following sense.  If one chooses
particular ``thresholds''---more accurately called ``switching
points''---to change the number of active flavors, then the evolution
kernels change at the thresholds.  Moreover, the matching conditions
at the thresholds can be thought of as corresponding delta-function
contributions to the kernels.

Some of the confusion in the literature can be traced to the
supposition that Wilson coefficients must be calculated with massless
quarks.  Indeed, many papers, for example
\cite{BMSvN,CFP,Ellis.et.al}, treat factorization as a question of
factoring out mass divergences in a massless theory.  Such methods
founder when the quarks have non-negligible masses, since then some of
the divergences are not literally present.  It should be noted that
the proof of factorization in \cite{CSS} does {\em not} assume that
quarks are massless (contrary to the assertion in \cite{RT1}); the
proof merely assumes that one is treating a limit in which the scale
of the hard scattering is much larger than all masses.

Another source of problems is
that many treatments of factorization \cite{BMSvN,CFP,Ellis.et.al}
take as their starting point an assertion
that hard cross sections are the
convolution of ``bare parton densities'' with unsubtracted
``partonic cross sections''.  Although this assertion is
widespread, it has no proof: it has the status of an unproved
conjecture.  Indeed it is not obvious that it is even true.
However, this bare parton conjecture is not necessary either to
the proof of the factorization theorem or to its use.

These problems with existing treatments, even without the treatment of
heavy quarks, provide motivation for providing much detail in the
proofs in this paper.  The proofs apply equally well in the absence of
heavy quarks.

The treatment in this paper will be based on the basic power counting
theorems derived by Libby and Sterman \cite{LS} and on the methods of
Curci, Furmanski and Petronzio \cite{CFP} for organizing sums of
generalized ladder graphs.  The treatment of heavy quarks uses the
methods of Collins, Wilczek and Zee (CWZ) \cite{CWZ}. The powerful
methods developed by Chetyrkin, Tkachov, and Gorishnii
\cite{perfect,CTG,AsyOp} for the operator product expansion with
mass effects are consistent with the CWZ scheme.

The outline of this paper is as follows: In Sec.\ \ref{sec:good}, I
explain the requirements that I consider necessary to impose on a
good treatment of mass effects.  Then, in Sec. \ref{sec:CWZ}, I review
the CWZ scheme for renormalization.  In that section, I also define a
consistent terminology of ``light'' and ``heavy'' quarks, and of
``partonic'' (or ``active'') and ``non-partonic'' quarks.  In Secs.\
\ref{sec:large.Q.basic} to \ref{sec:masses}, I prove factorization in
the case that there is one heavy quark and that $Q$ is at least as
large as the heavy quark mass; this is the case where the heavy quark
is active.  As an interlude in the formal proof, in Sec.\
\ref{sec:example}, I provide a mathematical example of the asymptotics
of certain integrals that mimic the behavior of the more complicated
integrals in Feynman graphs for QCD.  Then in Sec.\ \ref{sec:small.Q},
I prove factorization for the case that the heavy quark may be treated
as inactive. (``Non-partonic'' is a better term.)  The general case, that
there are several heavy quarks of various masses, forms a relatively
simple generalization of the preceding work, and is treated in Sec.\
\ref{sec:multiple}.  An account of the matching conditions and of the
evolution equations is given in Sec.\ \ref{sec:match.evolve}.  This is
followed by an account of the relation of the present scheme to the
schemes of other authors, in Sec.\ \ref{sec:comments}.  The
conclusions are in Sec.\ \ref{sec:concl}.  In the 
Appendix, I explain a certain mathematical
complication that appears in the middle of the proof.

\section{Requirements for a good factorization scheme}
\label{sec:good}

The overall aim of work such as ours is to represent interesting cross
sections (or other quantities) in terms of perturbatively calculable
quantities and a limited set of non-perturbative quantities that must
at present be obtained from experiment. A typical result is that for
deep-inelastic structure functions and other hard-scattering cross
sections we have factorization theorems: the leading large $Q$
behavior is a convolution of hard-scattering coefficients, which can
be perturbatively calculated, and of parton densities and/or
fragmentation functions. There are also evolution equations for the
parton densities, etc., for which the evolution kernels are
perturbatively calculable.

Although the factorization theorems are true in a general quantum
field theory, and not just in QCD, their particular utility in
QCD is caused by the asymptotic freedom of QCD.  Without the use
of factorization, perturbative calculations of typical scattering
amplitudes and cross sections involve integrals down to low
virtualities where the effective coupling is too large for
low-order perturbation theory to be valid. Factorization theorems
segregate the non-perturbative part of a cross section into a
limited number of experimentally measurable parton densities, etc.
Moreover, typical cross sections depend on several scales and
perturbative calculations typically have one or two logarithms of
ratios of scales for each loop.  Since the QCD coupling is not
very small, the logarithms can ruin the accuracy of practical
calculations. By working with quantities that each depend on a
single scale, one avoids this loss of accuracy.

For the purposes of this section, we will let $Q$ be a (large)
scale defining the kinematics of the hard-scattering process
under discussion and we will let $M$ denote the mass of some
heavy quark.  A satisfactory treatment should satisfy the
following requirements:
\begin{enumerate}

\item The formalism should apply to all orders of perturbation
    theory and include arbitrarily non-leading logarithms.

\item Explicit definitions must be given of the non-perturbative
    quantities, as matrix elements of operators.

\item The formalism is to be applicable to all the cases
    $Q \gg M$, $Q \sim M$ and $Q \ll M$, and the errors are suppressed
    by a power of $\Lambda /Q$.

\item Multiple heavy quarks should be treated without loss of
    accuracy no matter whether the ratios of the masses are large
    or not.

\end{enumerate}
The results in this paper will also satisfy some other
requirements which are more matters of convenience than absolute
principles:
\begin{enumerate}

\item When a quark mass is large enough for decoupling to apply,
    calculations should exhibit manifest decoupling.  That is,
    they should reduce to calculations in a standard scheme
    (e.g., \MSbar) in the theory with the heavy quarks omitted,
    and with no need to adjust the numerical values of the
    coupling.

\item The scheme should reduce to a standard scheme (e.g., \MSbar)
    when the masses are much less than $Q$.  We will in fact use
    the \MSbar{} scheme, so that standard hard-scattering
    calculations can be used unchanged in the case that masses
    can be neglected.

\item The previous two requirements apply to both factorization
    and to the coupling $\alpha _{s}$.

\item The evolution equations for the parton densities, etc., should
    be homogeneous.  That is, they should be of the form of
    conventional DGLAP equations or renormalization group
    equations rather than of the form of Callan--Symanzik
    equations \cite{Callan.Symanzik}.  (The solutions of
    Callan--Symanzik equations need an extra level of approximation
    to make them useful for calculations.)

\end{enumerate}

\section{CWZ scheme}
\label{sec:CWZ}

The short-distance coefficient functions are almost completely
determined once one has specified a scheme for defining the
parton densities --- in fact a scheme for renormalizing the
ultra-violet divergences in the coupling and in the parton
densities.  The scheme defined in this paper is in fact a
composite of a series of related schemes in the fashion proposed
by Collins, Wilczek and Zee (CWZ) \cite{CWZ}.

First, it is necessary to introduce some terminology whose consistent
use will aid our work.  Let us define a {\em ``light''} quark or gluon
to be one whose mass is of the order of $\Lambda $ or less, i.e.,
under about a GeV.  Similarly, let us define a {\em ``heavy''} quark to be one
whose mass is larger than a GeV or so, so that the the effective
coupling, $\alpha _{s}(M)$, at the scale of a heavy quark mass is in
the perturbative region.  With this definition, the charm, bottom and
top quarks are the heavy quarks.  We let $n_{l}$ be the number of
light quarks, and $n_{f}$ be the total number of quarks.  In our
present state of knowledge of QCD we have $n_{l}=3$ and $n_{f}=6$.

Each subscheme of the CWZ scheme is labeled by a number $n_{A}$,
which I will call the number of {\em ``active''} (or {\em
``partonic''}) quarks.  These are the $n_{A}$ lightest quarks.
All the remaining
quarks I call {\em ``non-partonic''}.  (It is also possible to call
them {\em ``inactive''}, but the term can be
misleading.)  In each subscheme:
\begin{enumerate}

\item
    Graphs that contain only active parton lines (i.e., gluons and
    active quarks) are renormalized by \MSbar{} counterterms, with the
    exception of the renormalization of the masses of heavy quarks.

\item
    Graphs all of whose external lines are active partons but which
    have internal non-partonic quark lines are renormalized by
    zero-momentum subtraction.

\item
    Heavy quark masses are defined as pole masses, as in the work of
    Smith, van Neerven and collaborators \cite{F2c,photo, BMSvN}. (We
    could also to choose to define heavy quark masses as \MSbar{}
    without changing the formalism.)

\item
    Other graphs with external non-partonic lines are renormalized by
    \MSbar{} counterterms.

\end{enumerate}
These definitions are applied to the renormalization of the
interaction and to the renormalization of the parton densities,
fragmentation functions, etc.

A consequence of the definitions is that we will talk about
``three-flavor'', ``four-flavor'', etc., definitions of the coupling
and parton densities (and fragmentation functions).  Use of such
a sequence of definitions is already common practice for the
coupling \cite{PDG}, and {\em identical} considerations apply to
the parton densities.  As a consequence it is meaningful to
specify numerical values of the coupling and of parton densities
only if the number of active flavors is specified.  There are
perturbatively calculable relations, or matching conditions,
between the values of these quantities with different numbers of
active flavors.

I will now list properties of this set of schemes that are
important for our purposes.  Their proofs are either in Ref.\
\cite{CWZ} or are later in this paper.
\begin{enumerate}

\item
    The scheme coincides with ordinary \MSbar{} when all partons
    are active,\footnote{
        Except that we have chosen to define heavy quark masses as
        pole masses.
    } i.e., $n_{A}=n_{f}$.

\item
    Manifest decoupling is obeyed.  If we have a process in which all
    external momentum scales are much less than the masses of the
    non-partonic quarks, then we can omit all graphs containing
    non-partonic quarks and only make a power-suppressed error.  In
    contrast, in a scheme that does not have manifest decoupling, we
    would have to adjust the numerical values of the couplings and of
    the parton densities.

\item
    Evolution equations for the densities of active partons and
    of the coupling $\alpha _{s}$ are {\em exactly} those of a pure
    \MSbar{} scheme in a theory with $n_{A}$ quark flavors.  This is
    a consequence of the mass-independence of UV counterterms in
    the \MSbar{} scheme, together with an application of the
    decoupling theorem \cite{AC,Witten}.

\item
    The relation between the subschemes is just a particular case of
    the relation between different renormalization schemes.  The
    matching conditions between the schemes with different numbers of
    active quarks are known to three loops for the coupling
    \cite{coupling-match} and to two loops for the parton densities
    \cite{pdf-match}.  The matching conditions between quantities in
    the subschemes with $N$ and $N+1$ active flavors involve no
    large logarithms of masses, provided that the
    renormalization/factorization scale $\mu $ is of order the mass of
    the $(N+1)$th quark.  (For example, we would choose $\mu $ to be of
    order the mass of the mass of the charm quark when we compute the
    relation between the three- and four-flavor schemes.)

\item
    In general, if one varies the physical scale $Q$ of some process
    (e.g., deep-inelastic scattering), one should vary the number of
    active quarks suitably.  Quarks of mass much less than $Q$ are to
    be active, while quarks of mass much larger than $Q$ should be
    non-partonic.  One has a choice for those flavors whose masses are
    close to $Q$, and I suspect a bias in favor of keeping quarks
    non-partonic will lead to more accurate calculations.

\item
    The light partons are always to be treated as active.

\end{enumerate}
It might be considered odd that in a region where $Q$ is of the order
of the mass of some heavy quark we have a choice as to whether to
treat the quark as active or not.  The freedom is entirely comparable
to the freedom to choose the precise value of the
renormalization/factorization scale.  Indeed the existence of a region
where the two subschemes have comparable accuracy is vital to the
success of a good treatment of heavy quarks, because it enables
reliable perturbative calculations to be made of the matching
conditions \cite{coupling-match,pdf-match} between the two
subschemes.

Commonly \cite{ACOT,RT2,PDG}, the scheme is implemented by choosing
what can be called ``matching'' and ``switching'' points to be equal
to the relevant heavy quark mass.  For example, in treating DIS with a
charm quark, one often sets the renormalization/factorization scale
$\mu $ to the kinematic variable $Q$.  Then one uses a 3-flavor
subscheme if $\mu < m_{c}$ and a 4-flavor subscheme if $\mu >
m_{c}$. One also chooses to evaluate the matching conditions between
the subschemes at $\mu =m_{c}$.  None of these choices is essential,
and any change gives a change in the physical predictions only because
of the errors due to the truncation of the perturbation series.  It is
probably only appropriate (i.e., suitable for fixed order perturbative
calculations) to use a 4-flavor subscheme if one is treating a
situation where the cross section is above the physical threshold for
charm production, which is at $Q = 2m_{c} \sqrt {x/(1-x)}$. Hence, if
$x$ is rather large, then it would be appropriate to use the 3-flavor
scheme even when $Q$ is substantially above $m_c$.

Note that there are three distinct mass scales referred to in the
previous paragraph: a matching point, a switching point and a
physical threshold.

Of course, one is free to disregard the CWZ scheme and use some other
scheme, provided that it provides complete definitions of the parton
densities and of the coupling.  However, this does not affect the
validity of the CWZ definitions.  The significance of the CWZ
definitions is that when all flavors are active, they are {\em
exactly} the \MSbar{} ones.

\section{Basics of factorization when $Q \protect\gtrsim M$}
\label{sec:large.Q.basic}

The principles of the proof can be best explained by first
considering the case that there is exactly one heavy quark, of
mass $M$.  There will be in effect two factorization theorems to
prove.  The first, whose treatment starts in this section, is
appropriate when the physical scale $Q$ of the hard scattering is
at least at large in magnitude as $M$.  In this case, it is
appropriate to treat the heavy quark as active: the factorization
theorem will include a term with a heavy quark density.

The second case, whose treatment starts in Sec.\ \ref{sec:small.Q}, is
appropriate when $Q \lesssim M$, and it treats the heavy quark as
non-partonic.  Then the factorization theorem has no term with a heavy
quark density, and all heavy-quark production is to be found (at
leading power) in the coefficient function.

As mentioned earlier, there is an overlap region, $Q \sim M$,
where both theorems are appropriate, i.e., they give comparable
accuracy in predictions based on finite-order calculations of
coefficient functions.

So in this section, we start the treatment of a factorization theorem
for deep-inelastic structure functions, given the assumption that $Q
\gtrsim M$. A single factorization formula will cover the case that
$Q$ is much bigger than the heavy quark mass, as well as the case that
$Q$ and the heavy quark mass are comparable, and the intermediate
region. Our notation for the photon momentum $q$, the hadron momentum
$p$, and for the Bjorken variable $x$ is standard.  As usual
$Q^{2}=-q^{2}>0$.  We will assume that quark masses are at most of
order $Q$.

When reading through the proof, it may be worth the reader's while to
refer ahead to Sec.\ \ref{sec:example}.  There, a simple mathematical
example is given of the kinds of integral under discussion, and it is
possible to see more easily the meaning of the formal manipulations in
the proof.

\subsection{Leading regions}
\label{sec:leading.region}

\begin{figure}
    \begin{center}
        \leavevmode
        \epsfxsize=3in
        \epsfbox{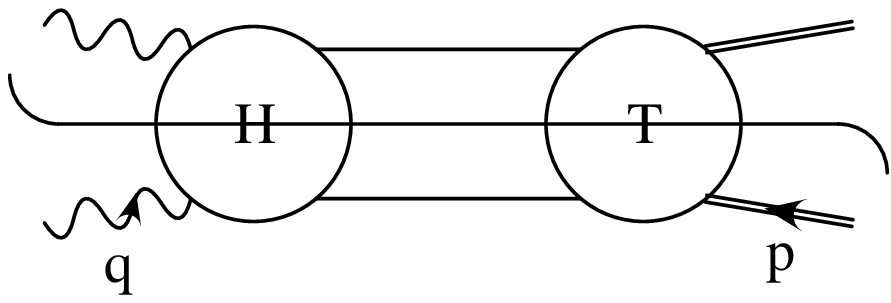}
    \end{center}
\caption{Regions for the leading power of structure functions
        have this structure.}
\label{fig:leading.region}
\end{figure}

In the Bjorken limit (large $Q$, fixed $x$), the leading power
behavior is given by the regions symbolized in Fig.\
\ref{fig:leading.region}, as was proved by Libby and Sterman
\cite{LS,Sterman.TASI95}.  In each region, there is what we call
a hard subgraph $H$, all of whose lines are effectively off shell
by order $Q^{2}$. It is to this subgraph that the virtual photons
couple. The rest of the graph has lines that are much lower in
virtuality and that are approximately collinear to the momentum
$p$ of the target. The latter part of the graph we will call the
target subgraph $T$.  We will give more quantitative
characterizations of the regions later.  (For example, we must
deal with the fact that there is a final-state cut, so that some
lines in $H$ are actually on shell instead of having virtuality
$Q^{2}$.)

Although one often does purely perturbative calculations
in which the target is a quark or gluon state, our treatment will
also apply to hadron targets.  In that case, suitable bound-state
wave functions will be incorporated in $T$.

A result of the power counting is that for a contribution to have the
leading power --- to be of ``leading twist'' --- the two subgraphs $H$
and $T$ must be connected to each other by two parton lines, one on
each side of the final-state cut. The set of decompositions into two
such subgraphs $H$ and $T$ is in one-to-one correspondence with the
set of leading regions.  There are two exceptions to this
correspondence.  The main exception is that if the heavy quark mass is
of order $Q$, then the $H$ and $T$ subgraphs are connected by light
parton lines, but not by heavy quark lines.  This exception arises
because the definition of the region implies that the lines joining
$H$ and $T$ have virtuality much less than $Q$, and this is not
possible if the lines are heavy quark lines of a mass comparable to
$Q$.  The second exception to the power-counting rules is that gluons
with scalar polarization can couple the $H$ and $T$ subgraphs without
a power-law penalty, at least in a covariant gauge: we will discuss
this issue in more detail later in the section.

We define the subgraph $T$ to include the full propagators of the
lines joining it to $H$, since these lines have momenta collinear
to the target. Hence the hard subgraph $H$ is
one-particle-irreducible (1PI) in these same lines.

In this and later figures, we have the initial state at the
bottom of the graph, and the hard subgraph to the left.  This
ensures that the orientation of the figures corresponds to the
equations we will write for convolutions of amplitudes.  For
example, we can write Fig.\ \ref{fig:leading.region} as $H\cdot T$.

Any region of loop-momentum space that cannot be characterized by
Fig.\ \ref{fig:leading.region} is suppressed by a power of $Q$.
Therefore the statement that the leading regions have the form
of Fig.\ \ref{fig:leading.region} is true to all orders in the
coupling and includes not just the leading logarithms but all
non-leading logarithms as well.

\begin{figure}
    \begin{center}
        \leavevmode
        \epsfxsize=3.5in
        \epsfbox{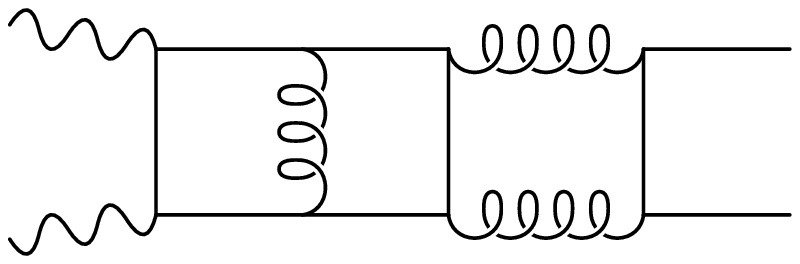}
    \end{center}
\caption{A graph with 4 decompositions of the form of Fig.\
        \protect\ref{fig:leading.region}.}
\label{fig:3rung.graph}
\end{figure}

A typical graph can have many different decompositions into hard and
target subgraphs.  For example, Fig.\ \ref{fig:3rung.graph} has four
such decompositions,\footnote{
   The one decomposition that may not be obvious is where $H$
   comprises the whole of the graph in Fig.\ \ref{fig:3rung.graph}
   with the exception of the right-most two external lines.  $T$ is
   then a trivial graph, in essence a factor of unity.
}
and hence four leading regions.  The possibility
of having more than one leading region is characteristic not only of
QCD, but of any renormalizable field theory, since adding extra lines
inside $H$ in a theory with a dimensionless coupling does not change
the counting of powers of $Q$.  It is the large multiplicity of
regions that results in many of the complications in the proof of
factorization.  In addition, it results in the logarithmic dependence
on $Q$ that is typical of higher order calculations in QCD.

\begin{figure}
    \begin{center}
        \leavevmode
        \epsfxsize=2.5in
        \epsfbox{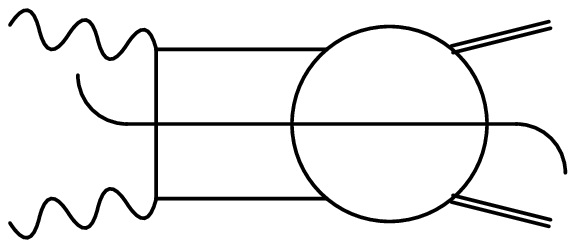}
    \end{center}
\caption{The handbag diagram that characterizes the only leading
        region in a super-renormalizable theory.}
\label{fig:handbag}
\end{figure}

\begin{figure}
    \begin{center}
        \leavevmode
        \epsfxsize=2.2in
        \epsfbox{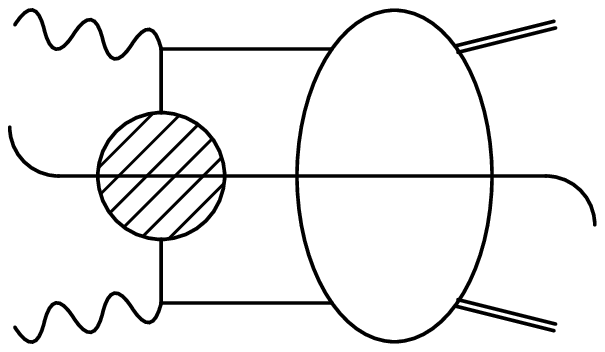}
    \end{center}
\caption{Handbag diagram with the final-state interactions that
    make the current quark jet.}
\label{fig:fsi}
\end{figure}

In contrast, super-renormalizable theories (e.g., QCD in less than
four space-time dimensions) have couplings with positive mass
dimension.  This implies that there is a single leading region.  It is
of the form of Fig.\ \ref{fig:leading.region}, but with the smallest
possible graph for $H$.  That is, the unique\footnote{ But see the
comments below concerning Fig.\ \ref{fig:fsi}.  } leading region has
the form of the handbag diagram, Fig.\ \ref{fig:handbag}. Although
super-renormalizable theories do not represent real strong-interaction
physics, experience in treating simple cases is useful in formulating
the factorization theorem.  Factorization, etc., for
super-renormalizable theories is equivalent to the set of results
obtained many years ago by Landshoff and Polkinghorne in the context of
their covariant parton model \cite{LP}.

Let us now list some technical complications that we will be able
to ignore, but that are treated in other papers \cite{LS,CS,CSS}
on factorization:
\begin{enumerate}

\item
    Although we have defined the target part $T$ to consists only of
    lines with collinear momenta, it may in fact contain some highly
    virtual lines.  These are confined to subgraphs that are
    ultra-violet divergent and just generate the usual UV divergences
    that are canceled by counterterms in the Lagrangian.  This
    complication does not affect our proofs, since none of the
    divergent subgraphs in QCD overlap between $T$ and $H$, and our
    proofs will treat $T$ as a black box.

\item 
\label{Unitarity.Sum}
    Although we treat the hard subgraph as being composed of lines all
    of which have large virtuality, this subgraph necessarily includes
    at least one final-state line. But after a sum over the possible
    final-state cuts, the hard subgraph is a discontinuity of a
    certain Green function. Then \cite{LS} the whole graph can be
    represented as a contour integral over a Green function in which
    all the lines in $H$ are off shell by order $Q^{2}$. Thus $H$ can
    indeed be treated as if its lines are all far off shell.  In
    particular, light-quark masses can legitimately be neglected
    compared to $Q$.  A simple example is given by a
    super-renormalizable theory.  Graphs with cut and uncut propagator
    corrections, Fig.\ \ref{fig:fsi}, to the handbag diagram have the
    same power law in $Q$ as the simple handbag diagram. Such graphs
    generate the correct final-state hadrons for the current-quark
    jet.  After a sum over cuts, all such corrections cancel at the
    leading power of $Q$, and the structure function is correctly
    given by the lowest order handbag Fig.\ \ref{fig:handbag}.

\item
    Soft gluons can connect the different final-state jets, and can
    connect the final-state jets to the target subgraph.  After a sum
    over final-state cuts these contributions cancel.  This
    complication is only present in a theory with elementary vector
    fields, e.g., QCD. A cancellation can be proved, and for the
    purposes of this paper, we may assume that no complications result
    from the implementation of the cancellation of soft gluons.  In
    more general processes, like the Drell-Yan process, the issue of
    soft-gluon cancellation is much more difficult \cite{CS,CSS}.

\item
    In a general gauge, there can be extra collinear gluon lines
    connecting $T$ and $H$.  Such gluons only contribute to the
    leading power if they have scalar polarization.  However, if a
    suitable ``physical'' gauge is used (e.g., axial gauge with a gauge
    fixing vector proportional to $q$), such contributions are not
    present \cite{LS}.  There are some subtleties associated with the
    use of such a gauge. For example, the analysis of the leading
    regions in Refs.\ \cite{LS,Sterman.TASI95} relies critically on
    Landau's analysis of the singularities associated with the
    denominators of Feynman propagators.  But physical gauges
    introduce extra unphysical singularities --- the physical gauges
    are not as physical as one often supposes.  For the purposes of
    this paper it is sufficient to ignore this complication, or to
    assume that the appropriate light-like gauge is being used.

\item
    The same phenomenon (in a covariant gauge) leads to what I term
    ``super-leading'' contributions, when $H$ and $T$ are joined only by
    gluons that have scalar polarizations.  It can be shown
    \cite{scalar-polarization} that the super-leading contributions
    cancel after a sum over a ``gauge-invariant set'' of graphs for $H$,
    and that \cite{CSS,scalar-polarization} the sum over attachments
    of scalar gluons to the hard part gives the correct
    gauge-invariant form of the parton densities, with a path-ordered
    exponential of the gluon field joining the two main parton
    vertices.

\end{enumerate}

\subsection{Relation of leading regions to mass singularities}

To characterize the regions of momenta that Fig.\
\ref{fig:leading.region} depicts, it is convenient to use
light-front coordinates, where we write a 4-vector $V$ as $V^{\mu } =
(V^{+},V^{-},{\bf V}_{T})$ with $V^{\pm } = (V^{0} \pm  V^{z}) / \sqrt 2$.
Then we choose a
coordinate frame such that
\begin{eqnarray}
   p^{\mu } &=& \left( p^{+}, \frac {m_{p}^{2}}{2p^{+}}, {\bf 0}_{T} \right) ,
\nonumber\\
   q^{\mu } &\approx & \left( -xp^{+}, \frac {Q^{2}}{2xp^{+}},
                {\bf 0}_{T} \right) .
\label{momenta}
\end{eqnarray}
The approximation in the definition of $q$ represents the neglect
of power suppressed terms, given that $x$ is normally defined as
$Q^{2}/2p\cdot q$.

To exhibit the counting of powers of $Q$ in its simplest form, we
will choose to boost the frame in the $z$ direction until
$p^{+}$ is of order $Q$.  Then regions of momentum corresponding to
the hard and target subgraphs are defined by saying that, for a
momentum $k^{\mu }$,
\begin{itemize}

\item $k$ is in $H$ if $k^{-}$ is of order $Q$.

\item $k$ is in $T$ if
    $k^{\mu } = \left( O(Q), o(Q), o(Q) \right)$, i.e.,
    $k^{+}$ is of order $Q$, while $k^{-}$ and $k_{T}$ are much smaller
    than $Q$, as is appropriate for a momentum collinear to the
    incoming hadron.\footnote{
        We use the mathematicians' big $O$ and little $o$
        notation:
        \\
        $A = O(Q)$ means that $A$ is of order $Q$ in the
        limit $Q\to \infty $.
        \\
        $A = o(Q)$ means that $A/Q \to  0$ in the limit $Q\to \infty $.
    }

\end{itemize}
After a sum over final-state cuts, the interactions that
hadronize the jets in the hard subgraph cancel
\cite{LS,Sterman.TASI95}, and then we may treat the lines in $H$
as if they are all off shell by order $Q^{2}$.

The gauge we are using is the light-cone gauge $A^{+}=0$. In this
gauge, regions with extra gluons joining the target and hard
subgraphs in Fig.\ \ref{fig:leading.region} are power suppressed.

Much of the literature treats factorization in terms of mass
singularities.  To see the relation to our treatment,
suppose that we were to take a limit of the structure
function in which all light quarks and all external lines are
massless. The target momentum would become light-like: $p^{\mu } \to
(p^{+},0,{\bf 0}_{T})$, so that there would be collinear and infra-red
divergences.  The infra-red divergences cancel after a sum over the
different possible graphs and final-state cuts at a given order of
perturbation theory, leaving only the collinear divergences
associated with the target.  These occur \cite{Sterman.TASI95} at
momentum configurations symbolized again by Fig.\
\ref{fig:leading.region}, but where momenta in $T$ are exactly
proportional to the target momentum, i.e., they are of the form
$k^{\mu } = (k^{+},0,{\bf 0}_{T})$.  There is an exact correspondence between
the leading regions (for any $m$) and the location of the
singularities for $m=0$: the leading regions are just
neighborhoods of the positions of the singularities. Moreover the
counting of powers of $Q$ corresponds to the degree of divergence
of the singularities.

However, in the true theory there need not be any actual divergences.
For example, in a non-QCD model we could endow all the particle with
masses, and our proof of factorization would remain correct.  In QCD
there are divergences that are associated with the necessary
masslessness of the gluon, but only if we make perturbative
calculations with on-shell external gluons or quarks.  In the real
world, these divergences are cut off by the non-perturbative effects
of confinement.  All the real particles of QCD are massive.  The
singularities in the massless limit merely provide a convenient tool
for {\em classifying} regions of momentum space.

\subsection{Elementary treatment of factorization}
\label{sec:basics}

The factorization theorem can easily be motivated from Fig.\
\ref{fig:leading.region}, as we will now show.  We will
construct an approximation to a proof of the theorem that will
introduce a number of useful ideas.  The proof will be exactly
correct in a super-renormalizable theory, where the single
important region is given in Fig.\ \ref{fig:handbag}.  In that
case the proof is equivalent to the argument given by Landshoff
and Polkinghorne for the parton model \cite{LP}.  The greater
detail given in the present paper will enable us to make precise
operator definitions of parton densities.  In addition, we will
introduce some notations and auxiliary concepts that will be
useful in the full proof.

The hypothesis on which the approximate proof rests is an assumption
that important momenta can be classified as belonging to either a
region of hard momenta (that belong only in $H$) or a region of
momenta collinear to the initial hadron $p$ (that belong only in $T$).
We will need to assume (not quite correctly) that the momenta
collinear to the target have virtualities that are fixed when $Q$
becomes large, and more specifically that the orders of magnitude of
the components of a target momentum are $(Q, m^{2}/Q, m)$, where $m$
is a typical light hadron scale.

Given this hypothesis,\footnote{
   Incidentally, this hypothesis excludes heavy quarks from
   consideration at this level of treatment, an error which we will
   remedy later.
} each graph can be decomposed unambiguously
into a sum of terms of the form of Fig.\
\ref{fig:leading.region}.  Thus we can write
\begin{eqnarray}
    F &=& \sum _{{\rm graphs}\ \Gamma } \Gamma
          + \mbox{non-leading power}
\nonumber\\
      &=& \sum _{{\rm graphs}\ \Gamma }
          \
          \sum _{{\rm regions}\ R}
             H(R) \cdot T(R)
          + \mbox{non-leading power} ,
\end{eqnarray}
where the summation over $\Gamma $ is restricted to those graphs that
are two-light-particle reducible in the $t$-channel and that therefore
have at least one decomposition of the form of Fig.\
\ref{fig:leading.region}. A region of such a graph is completely
defined by its hard and target subgraphs, so we can replace the
sum over graphs and regions by independent sums over graphs for
$H$ and $T$:
\begin{equation}
   F = H\cdot T + \mbox{non-leading power} .
\end{equation}
Here $H$ and $T$ are the sum over all possibilities for the $H$
and $T$ subgraphs in Fig.\ \ref{fig:leading.region}, with the
momenta being restricted to the appropriate regions. The symbol
$\cdot $ represents a convolution, the integral over the 4-momentum
linking $H$ and $T$ and a sum over the flavor, color and spin
indices of the lines joining the two subgraphs.  Thus we have
\begin{equation}
    H\cdot T = \sum _{i} \int  \frac {d^{4}k}{(2\pi )^{4}} \,
          H_{i}(q,k) \, T_{i}(k,p) .
\label{convolution}
\end{equation}
Recall that we defined $T$ to include the full propagators on the
two lines that connect it to $H$, so that $H$ is amputated in these
same two lines.

To get the factorization theorem, we use the observation that
some of the components of the loop momentum can be neglected in
$H$, and also that some of the components of the trace over spin
labels can be neglected. In the $H$ factor in Eq.\
(\ref{convolution}), we may neglect both $k^{-}$ and $k_{T}$, since all
the lines in $H$ are effectively off shell by order $Q^{2}$. This
results in an error that is suppressed by one or two powers of
$Q$. Thus we can approximate the structure function by:
\begin{equation}
   F = \int _{x}^{1} \frac {d\xi }{\xi }  H\left[q, (\xi p^{+},0,{\bf 0}_{T})
\right]
          \int  \frac {dk^{-} \, d^{2}{\bf k}_{T}}{(2\pi )^{4}} \xi p^{+}
T(k,p)
          + \mbox{non-leading power} .
\label{momentum.expansion}
\end{equation}
Here, to make contact with the standard usage in this subject, we have
written $k^{+}=\xi p^{+}$ and have changed variable from $k^{+}$ to
$\xi $.

In Eq.\ (\ref{momentum.expansion}) there is an implicit sum over the
spin indices and the flavor of the lines joining $T$ and $H$.  Suppose
the line is a quark.  Then we can decompose each of $H$ and $T$ into a
sum of Dirac $\gamma $ matrices.  The leading terms involve a $\gamma
^{-}$ in the target subgraph $T$ since that can be contracted with the
largest momentum components in $T$, which are the $+$ components.
Thus the most general form of the part of $T$ that gives the leading
power is a sum of terms proportional to $\gamma ^{-}$, $\gamma
^{-}\gamma _{5}$ and $\gamma ^{-}{\bf \gamma }_{T}$.

For the simple case of unpolarized scattering, only the $\gamma ^{-}$ term
contributes, and we can write\footnote{
    Generalization of the results to the polarized case results
    in purely notational complications, as regards the proof of
    factorization \cite{polfact}.
}
\begin{equation}
   F = \sum _{a} \int  \frac {d\xi }{\xi }  \tr H_{a} \gamma ^{-}
                   \int  \frac {dk^{-} \,d^{2}k_{T}}{(2\pi )^{4}} \xi p^{+}
\frac {1}{4} \tr \gamma ^{+} T_{a}
          + \mbox{gluon terms} + \mbox{non-leading power} ,
\label{collinear.approximation}
\end{equation}
with a similar decomposition being applied to the gluon term.
Here $a$ labels the different flavors of quark and antiquark.
(Note that in the usual applications, $H$ and $T$ are diagonal in
quark flavor and only a single flavor index is required, the same
for each of the lines joining $H$ and $T$.) A similar result
applies when $H$ and $T$ are joined by gluon lines.

It is convenient to represent this formula in a convolution
notation with the aid of a projection operator $Z$:
\begin{equation}
   F = H \cdot  Z \cdot  T + \mbox{non-leading power} .
\label{basic.factn}
\end{equation}
$Z$ represents the operation of setting $k_{T}=k^{-}=0$ for the momentum
of the external parton of the hard scattering and of picking out
the largest terms in the spin indices coupling the hard and
target subgraphs.  It is a sum of quark and gluon terms.  The
quark term is
\begin{equation}
   Z_{\alpha \alpha ';\beta \beta '}(k,l;\mbox{1st definition}) = \frac {1}{4}
\gamma ^{-}_{\alpha \alpha '} \gamma ^{+}_{\beta \beta '} \,
                   (2\pi )^{4}
                   \delta (k^{+}-l^{+}) \delta (k^{-}) \delta ^{(2)}({\bf
k}_{T}) .
\label{Z.def1}
\end{equation}
This and similar objects will be used repeatedly in our work. It is
readily verified that $Z$ is a projection, i.e.,
\begin{equation}
   Z^{2} = Z ,
\label{Z.is.projection}
\end{equation}
and hence, for example, $(1-Z)\cdot Z=0$.  The label ``1st definition''
in Eq.\ (\ref{Z.def1}) indicates that a modified definition, which we
will now give, is superior.

In fact, the above definition of the projector $Z$ is suitable for
massless quarks.  Its use in Eq.\ (\ref{basic.factn}) remains valid
when the quarks in $H$ have non-zero mass, but it is not perfectly
convenient for practical calculations.\footnote{
   Observe that in conventional treatments of factorization, it is
   normal to set quark masses to zero in the hard scattering.
   Precisely because we wish to treat heavy quarks, we do not at this
   point choose to set quark masses to zero.
}  For example, calculations of the short-distance
coefficient functions do not satisfy exact gauge invariance, because
the external lines of $H$ are off shell.  Therefore it is convenient
to replace Eq.\ (\ref{Z.def1}) by a definition in which the external
quarks of $H$ are put on shell. This involves replacing $k$ by an
on-shell momentum
\begin{equation}
    \hat k^{\mu } = (\xi p^{+},m^{2}/2\xi p^{+},{\bf 0}_{T}) ,
\label{k.hat}
\end{equation}
and using the Dirac matrix for on-shell wave functions:
\begin{equation}
   Z_{\alpha \alpha ';\beta \beta '}(k,l;\mbox{massive quark})
   = \frac {\hat k_{\mu }\gamma ^{\mu }_{\alpha \alpha '} + m}{4 k^{+}}
     \gamma ^{+}_{\beta \beta '}
     (2\pi )^{4}
     \delta (k^{+}-l^{+}) \delta (k^{-}-m^{2}/2k^{+}) \delta ^{(2)}({\bf
k}_{T}) .
\label{Z.def.massive}
\end{equation}
The resulting leading-power approximation to $F$ is
\begin{equation}
   H \cdot  Z \cdot  T =
   \sum _{a} \int  \frac {d\xi }{\xi }
       \tr H \frac{\hat k^{\mu }\gamma _{\mu } + m}{2}
   \int  \frac {dk^{-} \,d^{2}{\bf k}_{T}}{(2\pi )^{4}}
         \tr \frac {\gamma ^{+}}{2} T
   + \mbox{gluon terms} .
\end{equation}
Here $\hat k^{\mu }$ is the approximated momentum, Eq.\ (\ref{k.hat}).
Notice that although the external parton lines of $H$ are put
on-shell, this is not true of the corresponding external partons of
the target subgraph $T$; these are integrated over all values of
$k^{-}$ and $k_{T}$ in the collinear region of momentum.

The change in the definition of $Z$ for massive quarks does not
affect the factorization theorem (\ref{basic.factn}).  To
see this, observe that the change of definition only changes
small components of the momentum $k$ and of the $\gamma $ matrices
attached to $H$. Thus we have only made an error similar in size
to the power-suppressed error that we already induced by making an
approximation in the first place.  Also the algebraic property
$Z^{2}=Z$, which we will make frequent use of later, is unchanged.

Since the operation $Z$ projects out the integral over $k^{+}$, Eq.\
(\ref{basic.factn}) gives the structure function as a convolution
of a hard-scattering coefficient and parton densities:
\begin{equation}
   F = \hat F \otimes f + \mbox{non-leading power} .
\label{factorization}
\end{equation}
The symbol
$\otimes$ represents a convolution in the $\xi $ variable,\footnote{
    $\hat F \otimes f \equiv  \int d\xi /\xi \, \hat F(x,\xi ) \, f(\xi )$.
}
together with a sum over quark
flavors and over the gluon.  It will also include a sum over the
spin degrees of freedom if polarization-dependent effects are
being treated.

The parton densities can be expressed in their
usual form \cite{pdf.CS} as matrix elements of light-cone operators.
A quark density is then
\begin{equation}
   f(\xi ) = \int  \frac {dk^{-} \,d^{2}{\bf k}_{T}}{(2\pi )^{4}}
             \tr \frac {\gamma ^{+}}{2} T(k,p) .
\label{pdf.defn}
\end{equation}
Given that we obtained the factorization theorem by decomposing
momentum space into a hard region and a collinear region, the
integral in Eq.\ (\ref{pdf.defn}) is restricted to the collinear
region.  When we provide a more correct proof, we will remove the
restriction to collinear momenta, so that the definition of a
parton density is exactly as a matrix element of a bilocal
operator on the light-cone.

{}From the definition of $Z$, Eq.\ (\ref{Z.def.massive}), it then
follows that that the hard-scattering coefficient is computed
from $H$ by contracting with the Dirac matrices appropriate for
an external on-shell fermion, with a spin average:
\begin{equation}
   \hat F = \tr H \frac {\hat k^{\mu }\gamma _{\mu } + m}{2} .
\end{equation}
The factor of $1/2$ means that $\hat F$ has the normalization of a
spin-averaged cross section.

\subsection{Why the simple derivation does not work}

The above derivation of the factorization theorem would be valid if
one could use a fixed decomposition of momentum space into regions
appropriate for $H$ and $T$, at least up to power-suppressed
terms. This assumption is in fact true in a super-renormalizable
theory, and the above derivation then leads to the parton model. Only
the lowest order graph for $H$ gives a leading contribution in this
case, Fig.\ \ref{fig:handbag}. This kind of reasoning led Feynman
to formulate the parton model \cite{parton.model}.

Unfortunately the error estimates obtained from the above argument, in
a renormalizable theory, are of a relative size that we represent as
of order $(T/H)^{p}$.  Here we use $T$ to represent the largest
virtuality in the subgraph $T$, we use $H$ to represent the smallest
important virtuality in $H$, and $p$ is a fixed exponent.  In a
super-renormalizable theory there are leading power contributions only
when the virtualities in the subgraph $T$ are of order a hadronic mass
(squared), so we get an excellent error estimate.\footnote{ This fact
is established from the same power-counting rules that show that all
regions of the form of Fig.\ \ref{fig:leading.region} are leading in a
renormalizable theory.  }  But in renormalizable theories, including
QCD, there are logarithmic corrections that cover the whole range of
virtualities from a hadronic mass up to $Q$.  Thus the only simple
estimate of the errors is that they are of relative order unity, with
perhaps only a logarithmic suppression: the maximum virtuality in $T$
might only be a little less than the minimum virtuality in $H$. A more
powerful argument is needed to get a good proof of a theorem of the
form of Eq.\ (\ref{factorization}), with relative errors of order
$(\Lambda /Q)^{p}$, where $\Lambda $ denotes a typical hadronic
infra-red scale.

In addition, when we have heavy quarks, the proof does not give us a
factorization theorem that applies uniformly for any value of $Q$
larger than or of order of the quark mass.  If $Q$ is much larger than
$M$, the proof gives a factorization of just the same form as with
light quarks.  If $Q$ were of order $M$, then we would have to
restrict the lines joining $H$ and $T$ to be light partons, and then
to use the methods of Sec.\ \ref{sec:small.Q} below.  But the proof
would be unable to give an optimal error estimate in the intermediate
region.

\section{Proof of factorization when $Q \protect\gtrsim M$}
\label{sec:large.Q.proof}

Even with its defects, the reasoning in the previous section
contains a core of truth, which we will now use as the basis for
a correct proof.

Our aim is to prove
\begin{equation}
   F = \hat F \otimes f + \mbox{remainder} ,
\label{DIS.theorem}
\end{equation}
with the following properties:
\begin{enumerate}

\item
    The coefficient function $\hat F(x/\xi ,Q^{2},M^{2})$ is infra-red safe:
    it is dominated by virtualities of order $Q^{2}$.

\item
    The parton density $f$ is a renormalized matrix element of
    a light-cone operator.

\item
    The remainder is suppressed by a power of $\Lambda /Q$.

\item
    This suppression is uniform over the whole range $Q \gtrsim
    M$, so that, for example, there are no $O(M/Q)$ terms.

\end{enumerate}
This theorem looks just like the result (\ref{factorization}) we
tried to prove by elementary methods, except that the precise
definitions of the factors are different.

\subsection{Expansion in 2PI graphs}
\label{sec:2PI}

\begin{figure}
   \begin{eqnarray}
      \epscenterbox{1.1in}{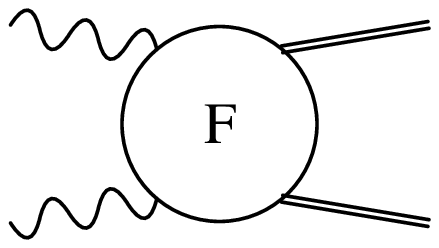}
   &~=~&
      \sum _{{\rm \# rungs} ~=~ 0}^{\infty }
           \epscenterbox{1.8in}{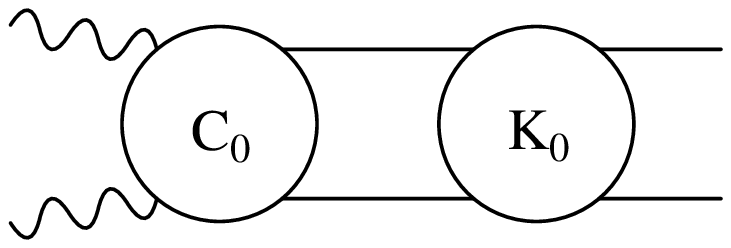}
           ~ \cdots ~
           \epscenterbox{1.8in}{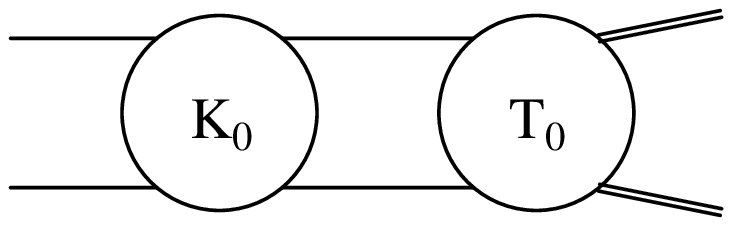}
   \nonumber\\*[0.2in]
      &&
      ~+~ \epscenterbox{1.1in}{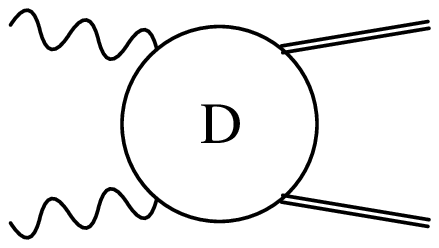}
   \nonumber
   \end{eqnarray}
\caption{Decomposition of structure function in terms of 2PI
        amplitudes.}
\label{fig:2PI}
\end{figure}

To utilize the result in Fig.\ \ref{fig:leading.region}, it is
convenient \cite{CFP} to decompose the structure function in
terms of two-particle irreducible amplitudes, Fig.\
\ref{fig:2PI}:
\begin{eqnarray}
    F &=& \sum _{n=0}^{\infty } C_{0} \cdot  (K_{0})^{n} \cdot  T_{0} + D
\nonumber\\
      &=& C_{0} \cdot  \frac {1}{1-K_{0}} \cdot  T_{0} + D .
\label{2PI}
\end{eqnarray}
The notations\footnote{
    The subscript zero in $C_{0}$, $K_{0}$ and $T_{0}$ is used because we
    will want to define some related but different objects later,
    with the same primary symbol, and we will in particular wish
    to reserve the unadorned symbol $C$ for the short-distance
    coefficient.
}
$C_{0}$ and $K_{0}$ are the same as in Ref.\
\onlinecite{CFP}.  Each of the amplitudes is two-particle
irreducible (2PI) in the horizontal channel (i.e., the
$t$ channel), except for the inclusion of full propagators
joining the amplitudes.
Thus $D$ is the 2PI part of the structure
function, while for the reducible graphs, $C_{0}$ is the 2PI
subgraph to which the currents couple, and $T_{0}$ is the 2PI
subgraph to which the target hadron couples.
Both $K_{0}$ and $T_{0}$ include full propagators\footnote{
    Strictly speaking, this means that to call the amplitudes 2PI
    is not quite correct.
}
on the left side, and consequently $C_{0}$ and $K_{0}$ are amputated on
the right, just as in Fig.\ \ref{fig:leading.region}. In
principle this is a non-perturbative decomposition. The
intermediate two-particle ``states in the $t$ channel'', between
the $C_{0}$, $K_{0}$, and $T_{0}$ factors, include all flavors of parton,
{\em including heavy quarks}.\footnote{
    In the case that the external hadrons are replaced by quarks
    or gluons, we will have $D=0$ and $T_{0}=1$.
}

\subsection{Construction of remainder}

It turns out to be convenient to first construct what will turn
out to be the remainder in Eq.\ (\ref{DIS.theorem}).  This is
defined by the following formula
\begin{eqnarray}
      r &=& \sum _{n=0}^{\infty } C_{0} \cdot  (1-Z) \cdot
            \left[ K_{0} (1-Z) \right]^{n} \cdot  T_{0} + D
\nonumber\\
        &=& C_{0} \cdot  \frac {1}{1 - (1-Z)K_{0}} \cdot  (1-Z) \cdot  T_{0} +
D
\nonumber\\
        &=& C_{0} \cdot  (1-Z) \cdot  \frac {1}{1 - K_{0}(1-Z)} \cdot  T_{0} +
D ,
\label{remainder}
\end{eqnarray}
with $Z$ being defined by Eq.\ (\ref{Z.def.massive}).  This formula is
obtained from the formula Eq.\ (\ref{2PI}) for the structure function
by inserting a factor $1-Z$ on each two-particle intermediate state in
the $t$ channel. This, as we will show, gives a power suppression.
The 2PI part, $D$, is non-leading since all the leading regions, Fig.\
\ref{fig:leading.region}, are associated with two-particle {\em
reducible} graphs.  The $1-Z$ factors may be considered as providing
subtractions that cancel all the leading regions.  That is, if we
start with the decomposition Eq.\ (\ref{2PI}) of the full structure
function and subtract off all leading contributions, then we end up
with Eq.\ (\ref{remainder}).

Once we know that $r$ as defined above is power suppressed, we will be
able to use the methods of linear algebra to construct a factorized
form for $F-r$.  This will be sufficient to give the factorization
theorem together with all the desired properties.

Now, leading contributions to the structure function come from
regions of the form of Fig.\ \ref{fig:leading.region}.  At the
boundary between the hard and target subgraphs, inserting a
factor of the operator $Z$ gives a good approximation.  Hence an
insertion of a factor $1-Z$ produces a power suppression.
Inserting a factor $1-Z$ at other places does not increase the
order of the magnitude of the graph.\footnote{Except that certain
    ultra-violet divergences may be introduced.  We will see
    later that are divergences when one separates the terms in
    Eq.\ (\ref{remainder}) with the $1$ and the $Z$ factors, but
    that there are no divergences in Eq.\ (\ref{remainder})
    itself.
}
Since we have put a factor $1-Z$ at every possible position of
boundary between hard and target subgraphs, we obtain a power
suppression for every term in Eq.\ (\ref{remainder}).

To be more concrete, suppose that we have a region of the form of Fig.\
\ref{fig:leading.region}. The insertion of a factor $1-Z$ at the
boundary between the region's hard subgraph and its target
subgraph gives a suppression by a factor of order
\begin{equation}
   \left( \frac {\mbox{highest virtuality in $T$}}{\mbox{lowest virtuality in
$H$}} \right) ^{p} ,
\label{suppresion.rung}
\end{equation}
as follows from the arguments in Sec.\ \ref{sec:basics}.

Furthermore, let us observe that in the left-most rung, closest
to the virtual photon, we have virtualities of order $Q^{2}$, while
in the right-most rung, closest to the target, we have
virtualities of order $\Lambda ^{2}$.  Within a given rung, the leading
power contribution comes where all the lines have comparable
virtualities, since leading power contributions only occur when
the boundaries of very different virtualities are as in Fig.\
\ref{fig:leading.region}.  Given that in Eq.\
(\ref{remainder}) we have a factor $1-Z$ between every 2PI rung,
there is a suppression whenever there is a strong decrease of
virtuality in going from one rung to its neighbor to the right.
Thus we find that Eq.\ (\ref{remainder}) has an overall
suppression of order
\begin{equation}
   \left( \frac {\Lambda }{Q} \right) ^{p} ,
\label{suppression.overall}
\end{equation}
when it is compared to the structure function itself
(\ref{2PI}).

This suppression of course gets degraded as one goes to higher
order for the rungs, since the lines within $K_{0}$ can have
somewhat different virtualities.  The larger a graph we have for
$K_{0}$, the wider the range of virtualities we can have without
meeting a significant suppression.

\subsection{Induced UV divergences}
\label{sec:UV}

The above argument shows that the quantity $r$, as defined by Eq.\
(\ref{remainder}), is power-suppressed in all the regions of momentum
space that are relevant for the structure function $F$.  However, the
existence of terms containing factors of $Z$ in Eq.\ (\ref{remainder})
entails some extra regions.  These regions have the potential of not
only being unsuppressed but also of giving UV divergences.

\begin{figure}
  \begin{displaymath}
      \epscenterbox{1in}{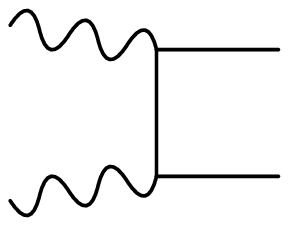}
      ~~~\otimes~~~
      \epscenterbox{2in}{kt.eps}
   \end{displaymath}
\caption{Second term of third line of Eq.\ (\protect\ref{r1}).}
\label{fig:f1-term}
\end{figure}

The lowest order non-trivial example is given by the $n=1$ term:
\begin{eqnarray}
    r_{1} &=& C_{0} \cdot  (1-Z) \cdot  K_{0} \cdot  (1-Z) T_{0}
\nonumber\\
       &=& C_{0} \cdot   K_{0} \cdot  (1-Z) T_{0}
       \ - \
           C_{0} \cdot  Z \cdot  K_{0} \cdot  (1-Z) T_{0}
\nonumber\\
       &=& C_{0} \cdot   K_{0} \cdot  (1-Z) T_{0}
        \ - \
            C_{0} \cdot  Z \cdot  K_{0} \cdot  T_{0}
        \ + \
            C_{0} \cdot  Z \cdot  K_{0} \cdot  Z \cdot  T_{0} .
\label{r1}
\end{eqnarray}
In the second term on the last line, the factor $Z \cdot K_{0} \cdot
T_{0}$ is a contribution to the matrix element of the bilocal operator
defining a parton density, Fig.\ \ref{fig:f1-term}.  There is a UV
divergence when the $k_{T}$ and $k^{-}$ in the loop(s) comprising the
operator vertex and the rung $K_{0}$ go to infinity.  The divergence
is in fact canceled by the last term in Eq.\ (\ref{r1}).  To see
this, observe that the two terms combine to give the second term on
the second line.  The $1-Z$ factor gives a power suppression of the
potentially divergent region, and the proof is the same as we used to
obtain the suppression proved in the previous subsection.  Look ahead
to Sec.\ \ref{sec:example} to see a concrete example illustrating the
above manipulations.

\begin{figure}
  \begin{displaymath}
      \epscenterbox{1in}{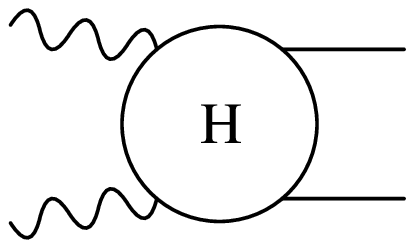}
      ~~~\otimes~~~
      \epscenterbox{1.8in}{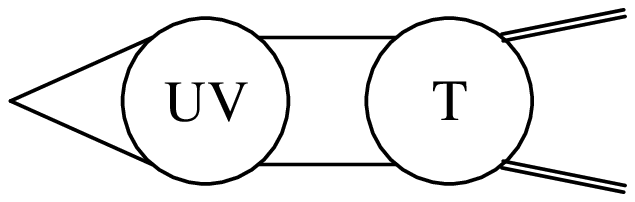}
   \end{displaymath}
\caption{Induced UV divergences in $r$ are in subgraphs of the
        form of $U$ in this diagram.}
\label{fig:rem-UV}
\end{figure}

A general proof of the cancellation of the induced UV divergences
immediately suggests itself.  The regions that give the possible
divergences arise from regions of the form shown in Fig.\
\ref{fig:rem-UV}.  There, the insertion of a $Z$ factor between two
rungs has given an operator vertex, through which can flow
ultra-violet momenta.  The proof of cancellation of the UV divergences
is simply that the $1-Z$ factors to the right suppress the regions
giving the UV divergences.

\subsection{Factorization}

We now derive a factorization formula for the structure function by
showing that $r$ is equal to the structure function minus the
factorized term in Eq.\ (\ref{DIS.theorem}).  Starting from Eqs.\
(\ref{2PI}) and (\ref{remainder}), we find
\begin{eqnarray}
    F - r &=& C_{0} \cdot  \left[
                       \frac {1}{1-K_{0}} - \frac {1}{1 - (1-Z)K_{0}} (1-Z)
                   \right]
                 \cdot  T_{0}
\nonumber\\
          &=& C_{0} \cdot  \frac {1}{1 - (1-Z)K_{0}} \cdot
                   \left[
                        1 - (1-Z)K_{0} - (1-Z)(1-K_{0})
                   \right]
                 \cdot  \frac {1}{1-K_{0}} \cdot  T_{0}
\nonumber\\
          &=& C_{0} \cdot  \frac {1}{1 - (1-Z)K_{0}} \cdot  Z
                 \cdot  \frac {1}{1-K_{0}} \cdot  T_{0} .
\label{main.proof}
\end{eqnarray}
This proof is very similar to some proofs in Refs.\ \cite{CFP} or
\cite{CSS-M}. It consists of some ordinary linear algebra, which is
valid since $Z$ and $K_{0}$ are just linear operators on the space of
4-momenta.  The form of the right-hand side of this equation is that
of the factorization theorem.  Aside from a normalization, the factor
$Z\cdot [1/(1-K_{0})] \cdot T_{0}$ is exactly the matrix element that
is a parton density, and then the remaining factor is the
short-distance coefficient function.

The only complication is the presence of UV divergences of the
form discussed in Sec.\ \ref{sec:UV}. There are divergences in
the parton density factor $Z\cdot  [1/(1-K_{0})] \cdot  T_{0}$ on the
right-hand side of Eq.\ (\ref{main.proof}).  There are also
divergences in the coefficient function $C_{0}\cdot [1/(1-(1-Z)K_{0})]$.  Of
course, these divergences cancel, since the left-hand side of
Eq.\ (\ref{main.proof}) is finite, as we have already proved.
For the moment, let us just apply any convenient UV regulator,
e.g., dimensional regularization.  We will show later how to
reorganize the right-hand side of Eq.\ (\ref{main.proof}) in
terms of UV finite quantities.

Given that there is a regulator, so that everything in Eq.\
(\ref{main.proof}) is well defined, we define a bare coefficient
function
\begin{equation}
         C_{B} = C_{0} \cdot  \frac {1}{1 - (1-Z)K_{0}} \cdot  Z ,
\label{coefficient.bare}
\end{equation}
and a bare\footnote{
    Our use of the terminology ``bare parton density'' has nothing
    in common with the usage in some other literature
    \cite{BMSvN,CFP,Ellis.et.al}.  In the present work, and in
    Ref.\ \cite{pdf.CS}, the word ``bare'' is used to denote a
    quantity that has ultra-violet divergences that have not been
    canceled by renormalization.  In
    \onlinecite{BMSvN,CFP,Ellis.et.al}, the word ``bare'' refers in
    some undefined sense to parton densities that are convoluted
    with unsubtracted partonic cross sections, and divergences in
    such a quantity are infra-red, not ultra-violet. See Sec.\
    \ref{sec:Z}, where we examine Zimmermann's methods, for a way
    of giving meaning to such formulas.
}
operator matrix element
\begin{equation}
         A_{B} = Z \cdot  \frac {1}{1-K_{0}} \cdot  T_{0} .
\label{operator.bare}
\end{equation}
This differs slightly in normalization from the parton densities
defined in Eq.\ (\ref{pdf.defn}), since $Z$ contains a $\frac {1}{2}
(\hat k^{\mu }\gamma _{\mu } + m)$ factor that we will ultimately put
in the coefficient function.  Other than that, the matrix element in
Eq.\ (\ref{operator.bare}) is the same as the parton density defined
in Eq.\ (\ref{pdf.defn}) when the momenta are unrestricted, which was
not the case in our derivation of Eq.\ (\ref{pdf.defn}).

{}From Eq.\ (\ref{main.proof}), together with the property that $r$ is
power suppressed, follows the factorization theorem
\begin{equation}
   F = C_{B} \otimes A_{B} + \mbox{non-leading power} .
\label{factorization.bare}
\end{equation}
Except for the subscripts, this equation has the same form as Eq.\
(\ref{factorization}). As in that equation, we have replaced the
symbol ``$\cdot $'' for convolution in 4-momentum by the symbol $\otimes$
for convolution in fractional $+$ momentum. The differences between
the two factorizations are that in Eq.\ (\ref{factorization.bare}) the
integrals defining the parton density and the coefficient are
unrestricted. Instead, the coefficient function, Eq.\
(\ref{coefficient.bare}), has factors of $1-Z$ placed between the 2PI
rungs. As we will see in an example in Sec.\ \ref{sec:example}, these
factors have the effect of making subtractions that prevent the double
counting of the different regions and of forcing the momenta in the
integrals for the coefficient function to be in the hard region of
virtuality of order $Q$. In contrast to this, the integrals in our
first approximation to a factorization theorem, Eq.\
(\ref{factorization}), are restricted to particular regions.
Moreover, for the new form of the factorization equation we have an
explicit estimate of the error, Eq.\ (\ref{suppression.overall}).

The bare matrix element $A_{B}$ is exactly a matrix element of a
particular bilocal light-cone operator.  This follows from the
fact that it is defined as an integral of the form of Eq.\
(\ref{pdf.defn}), with unrestricted integrals over $k^{-}$ and
${\bf k}_{T}$.

\section{Example}
\label{sec:example}

To understand the meaning of the above derivation, it is
convenient to examine a simple set of integrals that have the
same structure.

First, we observe that all the equations can be written as a sum
over powers in $K_{0}$, and that equations are true for each power
of $K_{0}$ separately.\footnote{
    Note that $K_{0}$ can be expanded in powers of the strong
    coupling $\alpha _{s}$, so that this expansion is related to the
    ordinary perturbation expansion.
}  Thus we can write the first few terms in the structure function
$C_{0}\frac {1}{1-K_{0}}T_{0}$ as
\begin{eqnarray}
    C_{0}\cdot  T_{0} &=&
    \begin{array}[t]{rccc}
         & [ C_{0} & \cdot  Z] \cdot  [Z \cdot  & T_{0} ]
    \\
       + &   C_{0} & \cdot   (1-Z)  \cdot  & T_{0}   ,
    \end{array}
\label{term0}
\\
    C_{0}\cdot  K_{0} \cdot  T_{0} &=&
    \begin{array}[t]{rccccc}
         & [ C_{0} & \cdot  Z ] \cdot  [ Z \cdot  & K_{0} &       \cdot
& T_{0} ]
    \\
       + & [ C_{0} &   \cdot  (1-Z) \cdot    & K_{0} & \cdot  Z ] \cdot  [ Z
\cdot  & T_{0} ]
    \\
       + &   C_{0} &   \cdot  (1-Z) \cdot    & K_{0} &   \cdot  (1-Z) \cdot
& T_{0}  ,
    \end{array}
\label{term1}
\\
    C_{0}\cdot  K_{0} \cdot  K_{0} \cdot  T_{0} &=&
    \begin{array}[t]{rccccccc}
         & [ C_{0} & \cdot  Z ] \cdot  [ Z \cdot  & K_{0} &       \cdot
& K_{0} &       \cdot        & T_{0} ]
    \\
       + & [ C_{0} &   \cdot  (1-Z) \cdot    & K_{0} & \cdot  Z ] \cdot  [ Z
\cdot  & K_{0} &       \cdot        & T_{0} ]
    \\
       + & [ C_{0} &   \cdot  (1-Z) \cdot    & K_{0} &   \cdot  (1-Z) \cdot
& K_{0} & \cdot  Z ] \cdot  [ Z \cdot  & T_{0} ]
    \\
       + &   C_{0} &   \cdot  (1-Z) \cdot    & K_{0} &   \cdot  (1-Z) \cdot
& K_{0} &   \cdot  (1-Z) \cdot    & T_{0}  .
    \end{array}
\label{term2}
\end{eqnarray}
The last term in each line is a power-suppressed and finite
remainder term, the contribution at the appropriate order in $K_{0}$
to the remainder $r$ defined in Eq.\ (\ref{remainder}). The other
terms are each a contribution to the coefficient function in Eq.\
(\ref{coefficient.bare}) times a contribution to the matrix
element in Eq.\ (\ref{operator.bare}). (I have used $Z^{2}=Z$ and
then the square-bracket notation to make this structure more manifest.)

\subsection{Model}

Now let us make a simple mathematical model that has all the
relevant structure.  We replace integrals over 4-dimensional
momenta by integrals over a 1-dimensional variable that runs
between $0$ and $\infty $, and we remove all labels for the flavor and
spin of the partons.  We also set the fully 2PI part $D$ of the
structure function to zero.  Then we define
\begin{eqnarray}
    C_{0}(k) &=& \frac {Q}{Q+k+m} ,
\nonumber\\
    K_{0}(k,l) &=& \frac {\alpha _{s}}{k+l+m} ,
\nonumber\\
    T_{0}(k) &=& \frac {1}{(k+m)^{2}} .
\label{analog.defns}
\end{eqnarray}
The motivations for these formulas are as follows:
\begin{itemize}

\item[$Q$] corresponds to the external photon momentum of
    deep-inelastic scattering, $m$ corresponds to a quark mass
    (heavy or light), and $k$ and $l$ correspond to the loop
    momenta coupling neighboring rungs in Eq.\ (\ref{2PI}).

\item[$C_{0}(k)$] is an analog of a lowest order graph for the hard
    part in Fig.\ \ref{fig:leading.region}.  In deep-inelastic
    scattering, it has a propagator that depends on a loop
    momentum $k$ plus a hard momentum $q$. This is modeled by the
    denominator $Q+k+m$.  The factor $Q$ in the numerator is
    inserted to provide a convenient normalization: $C_{0}\to 1$ as
    $Q\to \infty $.

\item[$K_{0}(k,l)$] is an analog of the lowest order graph for a
    rung.  The lowest order graph for $K_{0}$ in Eq.\ (\ref{2PI})
    has a dependence on a difference of external momenta, $k$ and
    $l$.  To make a simpler mathematical example, we have
    replaced $k-l$ by $k+l$.  To symbolize the analogy with a
    rung, we have put in a factor of the strong coupling $\alpha _{s}$,
    just as we would have for the lowest order rung in QCD. To ensure
    that the analogy is with a renormalizable theory, $K_{0}$ is defined
    in such a way that the coupling is dimensionless.

\item[$T_{0}(k)$] is given an extra power of $1/(k+m)$ compared with
    $K_{0}$.  Then it gives a finite result when integrated over all
    $k$, just as happens for $T_{0}$ in real QCD.  We could have
    used $T_{0}= 1/(k+p+m)^{2}$, with $p$ being like an external
    momentum.  But this would have been an irrelevant
    complication.

\end{itemize}
In each denominator in Eq.\ (\ref{analog.defns}), $m$ is meant to
be like a mass term. Just as in QCD we get a logarithmic
infra-red divergence when we have an integral over $K_{0}(k,l)$ with
respect to $k$, and we replace $l$ and $m$ by zero.

The mathematical structures we get are of the same form as in
QCD, but we will be able to present simple formulas.  For
example, there is no longitudinal $+$ component of momentum to
integrate over in the factorization formula.

To obtain examples of heavy quark physics, we can replace $m$ in
$C_{0}$ and/or some of the $K_{0}$'s by $M$.

\subsection{Lowest order}

The lowest-order term in the structure function $F$ is
\begin{equation}
   C_{0} \cdot  T_{0} = \int _{0}^{\infty } dk \frac {Q}{Q+k+m} \, \frac {
1}{(k+m)^{2}} .
\label{term.0}
\end{equation}
When $Q \to  \infty $, $k$ remains finite, and the asymptote is
\begin{equation}
   C_{0} \cdot  T_{0} \to  \int _{0}^{\infty } dk \frac {1}{(k+m)^{2}} .
\label{term.0.asy}
\end{equation}
Up to power suppressed factors, this is just the lowest order
coefficient function $C_{0} \cdot  Z$ times the lowest order matrix
element $Z \cdot  T_{0}$:
\begin{equation}
   C_{0} \cdot  Z \cdot  T_{0} = \frac {Q}{Q+m} \int _{0}^{\infty } dk \frac {
1}{(k+m)^{2}} .
\label{term.0.1}
\end{equation}
Here the operator $Z(k,l)$ is just $\delta (k)$.  That is, we get
$C_{0}\cdot Z\cdot T_{0}$ from $C_{0}\cdot T_{0}$ by setting $k=0$ in the
$C_{0}$ factor.

If we take $Q \to  \infty $ with $m$ fixed, the leading power behavior is
obtained by setting $m=0$ in the coefficient function: $Q/(Q+m) \to 1$.

\subsection{NLO term}

The next order term is
\begin{equation}
    C_{0} \cdot  K_{0} \cdot  T_{0} = \int _{0}^{\infty } dk \, \int
_{0}^{\infty } dl
                   \frac {Q}{Q+k+m} \, \frac {\alpha _{s}}{k+l+m} \, \frac {
1}{(l+m)^{2}} .
\label{term.1}
\end{equation}
There are two simple regions that give a leading power
$Q^{0}$: (a) $k$ and $l$ of order $m$, and (b) $k$ of order $Q$ with
$l$ of order $m$.  In addition the region $Q \gg k \gg l \sim m$
interpolates between the two simple regions and gives a
logarithmically enhanced contribution of order $\ln Q$.  This last
region gives the leading logarithm approximation.
It can be checked that the leading power contributions are all from
the region where $l \sim m$.

To derive the factorization formula expanded to order $K_{0}$, we
decompose $C_{0} \cdot K_{0}$ as follows:
\begin{eqnarray}
    C_{0} \cdot  K_{0} \cdot  T_{0} &=&
       ~~ C_{0} \cdot  Z \cdot  K_{0} \cdot T_{0}
\nonumber\\
     &&
       \, + \, C_{0} \cdot  (1-Z) \cdot  K_{0} \cdot  Z \cdot T_{0}
\nonumber\\
     &&
       \, + \, C_{0} \cdot  (1-Z) \cdot  K_{0} \cdot  (1-Z) \cdot T_{0},
\label{example.term1}
\end{eqnarray}
just as in Eq.\ (\ref{term1}).
We can explain the right-hand side of this equation as being
obtained by a series of successively improved approximations for the
leading behavior as $Q \to \infty $..

The first term on the right is the lowest-order coefficient times
the one-loop matrix element:
\begin{equation}
    C_{0} \cdot  Z \cdot  K_{0} \cdot  T_{0} =
                   \frac {Q}{Q+m} \int _{0}^{\infty } dk \, \int _{0}^{\infty }
dl
                   \, \frac {\alpha _{s}}{k+l+m} \, \frac {1}{(l+m)^{2}} .
\label{example.c0.m1}
\end{equation}
It gives a good approximation to the original integral Eq.\
(\ref{term.1}) in the region where $k$ and $l$ are of order $m$.
Its accuracy gets worse as $k$ increases.  Furthermore, we have
an ultra-violet divergence when $k \to  \infty $, since the
extra convergence at large $k$ given by the $Q/(Q+k+m)$ factor in
(\ref{term.1}) is removed by the approximation. In the real
factorization theorem in field theory, the divergence is the
normal UV divergence associated with the insertion of the vertex
for a composite operator (such as $\bar \psi  \gamma ^{\mu } \psi $).  To
define the
integral in Eq.\ (\ref{example.c0.m1}) we must implicitly apply
an ultra-violet regulator.  The regulator can be removed if we
apply suitable renormalization, as we will show in Sec.\
\ref{sec:example.renormalization}.

The poor approximation as $k$ increases towards $Q$ is remedied
by the second term in Eq.\ (\ref{example.term1}),
the one-loop coefficient times the lowest-order matrix element:
\begin{equation}
    C_{0} \cdot  (1-Z) \cdot  K_{0} \cdot  Z \cdot  T_{0} =
                   \int _{0}^{\infty } dk \, \int _{0}^{\infty } dl
                   \left( \frac {Q}{Q+k+m} - \frac {Q}{Q+m} \right)
                   \frac {\alpha _{s}}{k+m} \, \frac {1}{(l+m)^{2}} .
\label{example.c1.m0}
\end{equation}
This can be thought of as a term $C_{0} \cdot  K_{0} \cdot  Z \cdot  T_{0}$,
which gives
a good approximation when $k \sim Q$, together with a subtraction
term $- C_{0} \cdot  Z \cdot  K_{0} \cdot  Z \cdot  T_{0}$, which prevents
double counting
from the previous term, Eq.\ (\ref{example.c0.m1}).  The subtraction
term suppresses the contribution to Eq.\
(\ref{example.c1.m0}) of the infra-red region $k \ll Q$, so that
the one-loop contribution to the bare coefficient function
\begin{equation}
                   \int _{0}^{\infty } dk \,
                   \left( \frac {Q}{Q+k+m} - \frac {Q}{Q+m} \right)
                   \frac {\alpha _{s}}{k+m} ,
\label{example.c1.bare}
\end{equation}
has no IR divergence in the massless limit.  This term also has a UV
divergence equal and opposite to that in Eq.\ (\ref{example.c0.m1}),
so that the sum of the two terms is UV finite.

The structure of the subtraction terms is exactly the same as in
the work of Aivazis {\it et al.}\ \cite{ACOT} on calculations of
coefficient functions for heavy quark processes.  To get a more
exact analogy to that work, one could change $C_{0}$ to $Q/(Q+k+M)$,
i.e., one could replace the light quark mass in $C_{0}$ by a heavy
quark mass.  This mimics the effect of a heavy quark loop at the
left-hand end of the diagram (confined to $C_{0}$).  It is left to
the reader to check that all the statements we make about the
asymptotic behavior remain true in this heavy quark example,
provided only that $Q$ is large compared to the {\em light} quark
mass $m$, and that $Q$ is roughly at least as large as the heavy
quark mass $M$.  That is the remainder is suppressed by $m/Q$
rather than just $M/Q$.

\subsection{NLO: remainder}

The third term on the right of Eq.\ (\ref{example.term1}) is the
remainder.  It is simply the left-hand side minus the first two
terms. The fact that the sum of the first two terms gives the
full leading power, complete with its logarithm, is demonstrated
by showing that the remainder,
\begin{eqnarray}
    C_{0} \cdot  (1-Z) \cdot  K_{0} \cdot  (1-Z) \cdot  T_{0} &=&
                   \int _{0}^{\infty } dk \, \int _{0}^{\infty } dl
                   \left( \frac {Q}{Q+k+m} - \frac {Q}{Q+m} \right)
\nonumber\\
       &&~~~~~~~~
                   \left( \frac {\alpha _{s}}{k+l+m} - \frac {\alpha _{s}}{k+m}
\right)
                   \frac {1}{(l+m)^{2}} ,
\label{example.rem1}
\end{eqnarray}
is power suppressed.  To see this, we observe that the
potentially leading contributions, when $k \lesssim Q$ and
$l \sim m$ are canceled by the subtractions.\footnote{
    $Q \gtrsim k$ includes the regions $k \sim Q$ and $k \ll Q$.
}
There is a possible UV divergence as $k \to  \infty $, but this is
canceled by the subtraction in the second factor.  This
subtraction suppresses the region $k \gg l$, and it is as
effective at suppressing the region for the ultra-violet
divergence, viz.\ $k \to  \infty $, as it is at suppressing the original
region it was designed to handle, $k \sim Q$.

\subsection{NLO: renormalization}
\label{sec:example.renormalization}

Next, we perform renormalization in the two terms contributing to
the leading power.  We can remove the UV divergence in each term
separately by adding suitable counterterms; in the factorization
theorem this would amount to defining renormalized composite
operators, a procedure we will implement in Secs.\
\ref{sec:op.ren}--\ref{sec:factn.ren}. A convenient method of
constructing counterterms is subtraction of the asymptote
\cite{sub.asy}.  So we can define the lowest-order coefficient
times the renormalized two-loop matrix element to be
\begin{equation}
    R \left( C_{0} \cdot  Z \cdot  K_{0} \cdot  T_{0} \right) =
                   \frac {Q}{Q+m} \int _{0}^{\infty } dk \, \int _{0}^{\infty }
dl
                    \left( \frac {\alpha _{s}}{k+l+m} - \frac {\alpha _{s}
\theta (k>\mu )}{k} \right)
                     \frac {1}{(l+m)^{2}} .
\label{example.c0.m1.ren}
\end{equation}
In field theory, a sensible counterterm to a subgraph is a polynomial
in the external momenta of the subgraph.  If we use minimal
subtraction, the counterterm is also polynomial in masses.  The degree
of the polynomial is equal to the degree of divergence.  In our toy
example, this means that the counterterm has to be independent of $l$
and $m$.  The counterterm $\alpha _{s} \theta (k>\mu ) /k$ does indeed
satisfy this criterion.  The $\theta $ function is needed to prevent
there from begin an infra-red divergence in the counterterm, and the
arbitrary parameter $\mu $ has the function of a
renormalization/factorization scale, just as in conventional minimal
subtraction.

It now follows that the renormalized one-loop coefficient function is
\begin{equation}
  R\left( C_{0} \cdot  (1-Z) \cdot  K_{0} \cdot  Z \right)
=   \int _{0}^{\infty } dk
    \left[
        \left( \frac {Q}{Q+k+m} - \frac {Q}{Q+m} \right)
              \frac {\alpha _{s}}{k+m}
        \, +\,  \frac {Q}{Q+m}
          \frac {\alpha _{s} \theta (k>\mu )}{k}
     \right],
\label{example.c1.ren}
\end{equation}
which is multiplied by the one-loop matrix element $\int _{0}^{\infty  } dl
/(l+m)^{2}$. The counterterms in the above two terms are equal
and opposite, so that the sum of the two renormalized contributions to
the leading power is the same as the sum of the bare terms.  Notice
that if we choose the factorization scale $\mu $ to be of order $Q$,
then the integral in the one-loop coefficient function is dominated by
$k$ of order $Q$.

\subsection{Zero mass limit of coefficient function}

Finally, we observe that the coefficient function has a finite $m
\to  0$ limit.  The coefficient function is the sum of the lowest
order term $C_{0}\cdot Z=Q/(Q+m)$, the one-loop term Eq.\
(\ref{example.c1.ren}), and higher-order terms.
In a field theory, the existence of the zero mass limit
implies that the coefficient function is infra-red safe and is a
symptom of the perturbative computability of the coefficient
function in QCD when $Q$ is large.

For example, the massless limit of Eq.\ (\ref{example.c1.ren}) is
\begin{equation}
    \int _{0}^{\infty } dk
    \left[
        \left( \frac {Q}{Q+k} - 1 \right) \frac {\alpha _{s}}{k}
     \, + \, \frac {\alpha _{s} \theta (k>\mu )}{k}
     \right] .
\label{example.c1.ren.massless}
\end{equation}
The infra-red divergence (at $k=0$) in the term
$\int dk \frac {Q}{Q+k} \frac {\alpha _{s}}{k}$ is canceled by the
subtraction in the first term.  The subtraction is designed to
cancel the region where $k \ll Q$, and this includes the region of the
possible infra-red divergence.

One reason for emphasizing the zero mass limit is that calculations
become algorithmically much simpler, especially for the analytic
evaluation of Feynman graphs.  But our derivation shows that a
non-zero mass may be left in the calculation of the coefficient
functions, as would be appropriate if the mass is not sufficiently
small compared with $Q$.

\section{Use of renormalized parton densities}

We now return to the factorization theorem in field theory.

\subsection{Renormalization of operators}
\label{sec:op.ren}

To construct the final form of the factorization, we will
re-express the bare factorization theorem, Eq.\
(\ref{factorization.bare}), in terms of the matrix elements of
renormalized operators. These operators have no UV divergences,
unlike the bare operator matrix elements defined in Eq.\
(\ref{operator.bare}).

\begin{figure}
   \begin{displaymath}
      {\rm \large UV}\left(\epscenterbox{1.2in}{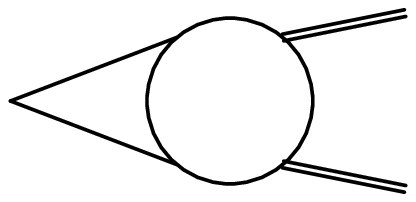}~~\right)
      ~=~
      \epscenterbox{2in}{uv.eps}
   \end{displaymath}
\caption{Regions of momentum integration that give the UV
         divergences in the operator matrix element defined by
         Eq.\ (\ref{operator.bare}).
}
\label{fig:UV.divergences}
\end{figure}

Now, the divergences come from regions of the form shown in Fig.\
\ref{fig:UV.divergences}.  This figure is very reminiscent of
Fig.\ \ref{fig:leading.region}, for the very good reason that the
derivation of the associated regions is essentially identical for
the two cases. We will choose to renormalize the divergences in
the \MSbar{} scheme using dimensional regularization.  As we will
see, the fact that the counterterms in this scheme are
mass independent will permit us to take the zero mass limit for
the coefficient function without encountering mass divergences
introduced by the renormalization counterterms. Minor changes to
the argument would permit the use of any other suitable scheme.

\begin{figure}
    \begin{center}
        \leavevmode
        \epsfxsize=2in
        \epsfbox{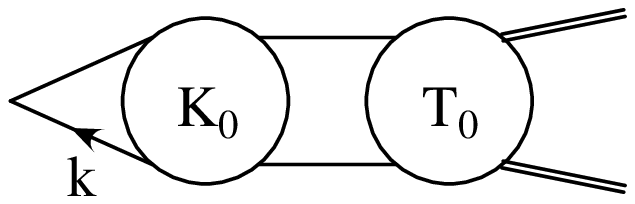}
    \end{center}
\caption{One-rung graph for the matrix element.}
\label{fig:one.rung}
\end{figure}

To see what to do, let us first expand the bare operator matrix
element, $A_{B}$, in powers of $K_{0}$:
\begin{eqnarray}
    A_{B} &=& Z \cdot  T_{0}
           ~+~ Z \cdot  K_{0} \cdot  T_{0}
           ~+~ Z \cdot  K_{0} \cdot  K_{0} \cdot  T_{0}
           ~+~ \cdots .
\end{eqnarray}
The first term is UV finite.  The second term has a divergence
when the loop momentum $k$ joining the operator vertex and $K_{0}$
(Fig.\ \ref{fig:one.rung}) goes to infinity.  It can be
renormalized by subtracting the pole part at $\epsilon =0$.  (We define
the number of space-time dimensions to be $4-\epsilon $.)  This gives a
result we symbolize as
\begin{eqnarray}
   R \left[ Z \cdot  K_{0} \cdot  T_{0} \right]
     &=& Z \cdot  K_{0} \cdot  T_{0}
         ~-~ \mbox{pole part} \left( Z \cdot  K_{0} \right) \cdot  T_{0}
\nonumber\\
     &=& Z \cdot  K_{0} \cdot
         \left( 1 - \polepart \right) \cdot  T_{0} .
\end{eqnarray}
Here $\polepart $ means to take the pole part of
everything to its left, with the usual modifications of the pole
part that define the \MSbar{} scheme.  Although we have used a
notation that suggests $\polepart $ is to be treated as
a linear operator, it does not\footnote{
    Compare the remarks of Curci, Furmanski and Petronzio below
    Eq.\ (2.25) of Ref.\ \onlinecite{CFP}, and see also the Appendix
    of the present paper.
}
in fact obey all the properties of linear operators, in
particular associativity.

\begin{figure}
    \begin{center}
        \leavevmode
        \epsfxsize=3in
        \epsfbox{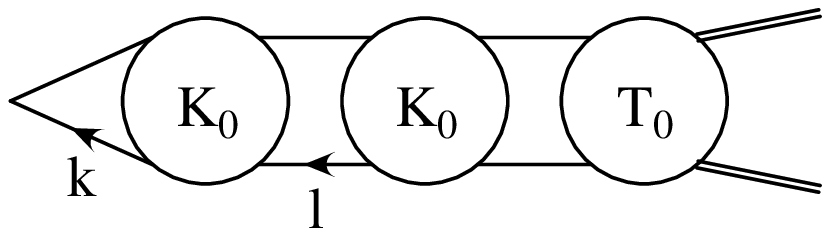}
    \end{center}
\caption{Two-rung graph for the matrix element.}
\label{fig:two.rung}
\end{figure}

Renormalization of graphs with two or more rungs is more
interesting.  For example the two-rung graphs, Fig.\
\ref{fig:two.rung}, have a sub-divergence as the left-most loop
momentum $k$ goes to infinity; this is exactly the same
divergence as in the one-rung graphs Fig.\ \ref{fig:one.rung}.
It must be canceled by the one-rung counterterm before we add in
the counterterm for the two-rung divergence, which occurs when
both the loop momenta, $k$ and $l$, go to infinity.  Note that
there will also be UV divergences inside each rung from
divergent self-energy and vertex graphs.  These are associated
with renormalization of the Lagrangian and are present
independently of the UV divergences that we are discussing now,
divergences that are due to the use of composite operators.  The
divergences associated with the interactions are canceled by the
usual collection of counterterms in the Lagrangian, so that $C_{0}$,
$K_{0}$ and $T_{0}$ are finite before we convolute them together.  This
implies, in particular, that the Green functions that define
these amplitudes are Green functions of {\em renormalized}
fields.

According to this procedure, the one-rung divergence in Fig.\
\ref{fig:two.rung} is canceled by a counterterm
\begin{equation}
     - Z \cdot  K_{0} \cdot  \polepart  \cdot  K_{0} \cdot  T_{0} ,
\end{equation}
and so the two-rung counterterm is
\begin{equation}
    - Z \cdot  K_{0} \cdot  \left( 1- \polepart \right)
      \cdot  K_{0} \cdot  \polepart  \cdot  T_{0} .
\label{two.rung.ct}
\end{equation}
The important point in the definition of $\polepart $
is that it must only be applied to quantities (to its left) that
are free of subdivergences.  To do otherwise would generate
counterterms that have non-polynomial dependence on the external
momenta and that can therefore not be interpreted in terms of
operator renormalization.
The renormalized value of the operator to two-rung order is
therefore
\begin{equation}
    Z \cdot  K_{0} \cdot
    \left( 1 - \polepart \right) \cdot  K_{0} \cdot
    \left( 1 - \polepart \right) \cdot  T_{0} .
\end{equation}

This pattern evidently generalizes.  To renormalize the operator
matrix element, we simply insert a factor of
$1-\polepart $ to the right of every  $K_{0}$ factor.
The result is that the renormalized matrix element is
\begin{eqnarray}
    A_{R} &=& \sum _{n=0}^{\infty }
    Z \cdot  \left[ K_{0} \cdot  \left( 1 - \polepart \right)
        \right] ^{n}
    \cdot  T_{0}
\nonumber\\
    &=&
    Z \cdot  \frac {1}{1 - K_{0} \cdot  \left( 1 - \polepart \right)}
    \cdot  T_{0} .
\label{operator.ren}
\end{eqnarray}
The structure here is very similar to our construction of the
remainder, Eq.\ (\ref{remainder}).  This is not surprising, since
in both cases we are cancelling contributions from a set of
regions of loop-momentum space that have very similar structures.

Given that $Z$ effectively represents the vertices for the
operators that define parton densities, Eq.\ (\ref{operator.ren})
is our definition of the parton densities, up to a trivial
normalization factor.

\subsection{Operator renormalization is multiplicative}

At first sight, the above manipulations give a rather arbitrary
definition of the renormalization of the operators and of the
parton densities.  In fact, as we will now show, they give a
definition in which the renormalized and bare parton densities
differ by a multiplicative factor, with the multiplication being
in the sense of convolution over fractional longitudinal
momentum. Therefore the only freedom is the usual
renormalization-group freedom to change the renormalization
scheme or to change the scale parameter(s) within a particular
scheme.

What enables these results to be proved is the fact that
renormalization counterterms are polynomial in the external
momenta of the subgraph to which they apply.  Thus the
counterterms can be interpreted as factors times operator
vertices.  (The same property is what enables renormalization of
the interaction to work.)  Moreover, the fact that the
divergences are logarithmic implies that the operator vertices
are just the ones defining the bare parton densities. These
properties can be summarized by the statement that multiplying
$\polepart $ on the right by $Z$ has no effect:
\begin{equation}
   X \cdot  \polepart  = X \cdot  \polepart  \cdot  Z .
\end{equation}
Here $X$ is any quantity which is free of subdivergences.

Now we can express the renormalized parton densities $A_{R}$ in
terms of the bare parton densities:
\begin{eqnarray}
    A_{R} &=&
        Z \cdot  \frac {1}{1 - K_{0} \cdot  \left( 1 - \polepart \right)}
        \cdot  T_{0}
\nonumber\\
     &=&
        Z \cdot  \frac {1}{1 - K_{0} \cdot  \left( 1 - \polepart \right)}
        (1-K_{0}) \cdot  \frac {1}{1-K_{0}}
        \cdot  T_{0}
\nonumber\\
    &=&
    Z \cdot  \frac {1}{1 - K_{0} \cdot  \left( 1 - \polepart \right)}
    \left( 1-K_{0} (1-\polepart )
           - K_{0} \polepart
    \right)
    \cdot  \frac {1}{1-K_{0}} \cdot  T_{0}
\nonumber\\
    &=&
    \left[ Z -
           Z \cdot  \frac {1}{1 - K_{0} \cdot  \left( 1 - \polepart \right)}
             \cdot  K_{0} \polepart
    \right]
    \cdot  Z \cdot  \frac {1}{1-K_{0}} \cdot  T_{0}
\nonumber\\
    &=& G \otimes A_{B} .
\label{operator.ren.proof}
\end{eqnarray}
In the next-to-last line, we have used $Z^{2}=Z$ and
$\polepart \cdot  Z = \polepart $, to write the
result in terms of an explicit factor times the bare operator
matrix element.  Then we observe that there is a factor $Z$ at
the left of the operator matrix element $Z \cdot  \frac {1}{1-K_{0}} \cdot
T_{0}$ and that
the integral coupling it to everything further to the left only
involves the $+$ component of momentum.  Thus the result has the
form of a convolution over longitudinal momentum fraction, for
which we use the symbol $\otimes$.

The factor
\begin{equation}
    G \equiv  Z -
        Z \cdot  \frac {1}{1 - K_{0} \cdot  \left( 1 - \polepart \right)}
             \cdot  K_{0} \polepart
\label{ren.factor}
\end{equation}
is the renormalization factor of the operator defining the parton
densities.  We can therefore write the renormalized parton
densities in terms of the unrenormalized ones:
\begin{equation}
   f^{R}_{i/p}(x) = \sum _{j} \int  \frac {d\xi }{\xi } 
   \, G_{ij} (\xi /x, \alpha _{s}, \epsilon ) \, f^{B}_{j/p}(\xi ) ,
\end{equation}
where we have now explicitly displayed the sum over parton
flavors and the integral over momentum fraction $\xi $.
Let us reiterate that the word ``bare'' is used in the sense of
``lacking UV renormalization'', and has no connection with
another common usage of the word in this context
\cite{BMSvN,CFP,Ellis.et.al}.  The renormalization factor starts
with a lowest order term which is effectively a unit operator:
\begin{equation}
   G_{ij} = \delta _{ij} \delta (\xi /x-1) + O(\alpha _{s}) .
\end{equation}

\subsection{Factorization with renormalized parton densities}
\label{sec:factn.ren}

Once we have seen that the renormalization of the operators is
multiplicative, we can write the factorization theorem Eq.\
(\ref{factorization.bare}) in terms of renormalization
quantities:
\begin{equation}
   F = C_{R} \otimes A_{R} + \mbox{remainder, $r$} ,
\label{factorization.ren}
\end{equation}
where the renormalized coefficient function is
\begin{equation}
   C_{R} = C_{B} \otimes G^{-1} ,
\label{coefficient.ren}
\end{equation}
with $G^{-1}$ being the inverse of the renormalization factor $G$ for
the parton densities $A_{R}$.  The inverse is with respect to
convolution in the longitudinal momentum fraction.

It is possible to derive a simple and very plausible, but wrong,
formula for the renormalized coefficient function.  The
derivation relies on using associativity for the pole part
operation.  We give the false derivation in the Appendix,
since it is instructive.

There does not appear to be a simple closed formula for the
renormalized coefficient function.  But there is
a convenient recursion relation that we will now derive.  It
corresponds to the actual
algorithms used to do real calculations.

The derivation starts from the fact that by our definition of $C_{R}$,
\begin{equation}
    C_{R} \otimes A_{R} = C_{B} \otimes A_{B} .
\label{bare.and.renormalized.factor}
\end{equation}
We simply expand all quantities in this in powers of $K_{0}$.  Since
we already know the
$n$th order terms for $C_{B}$, $A_{B}$, and $A_{R}$:
\begin{eqnarray}
    C_{B}^{(n)} &=& C_{0} \left[ (1-Z) K_{0} \right]^{n} Z ,
\nonumber\\
    A_{B}^{(n)} &=& Z K_{0}^{n} T_{0} ,
\nonumber\\
    A_{R}^{(n)} &=& Z \left[ K_{0} (1-\polepart) \right]^{n} T_{0} ,
\end{eqnarray}
we can obtain the expansion of $C_{R}$, which we write as
\begin{equation}
   C_{R} = \sum _{n=0}^{\infty } C_{R}^{(n)}.
\end{equation}
Our problem is to find an explicit formula for the term $C_{R}^{(n)}$,
given the lower order terms.

Expanding Eq.\ (\ref{factorization.ren}) to zeroth order in $K_{0}$, we
find
\begin{equation}
   C_{0} Z T_{0} = C_{R}^{(0)} Z T_{0} .
\end{equation}
This equation is true for any value of $T_{0}$, since factorization
applies for any initial state.  Hence we must have
$C_{R}^{(0)}=C_{0}Z$, the same as corresponding term in the bare
coefficient.

To first order, we have
\begin{equation}
   C_{B}^{(1)} A_{B}^{(0)} + C_{B}^{(0)} A_{B}^{(1)} =
   C_{R}^{(1)} A_{R}^{(0)} + C_{R}^{(0)} A_{R}^{(1)} ,
\end{equation}
which gives
\begin{eqnarray}
   C_{R}^{(1)} &=& C_{0} (1-Z) K_{0} Z 
     ~+~ \left[  C_{0} Z \right] \left[ (ZK_{0}) \polepart \right]
\nonumber\\
     &=& C_{0} K_{0} Z
         ~-~ C_{0} Z  \left[ Z K_{0} - (ZK_{0})\polepart \right] .
\label{first.order.coeff}
\end{eqnarray}
A convenient way of formulating this is to say that the
right-hand side is the structure function of an on-shell quark
(or gluon) minus the lower order term in the Wilson expansion of this
partonic structure function.  

Notice very carefully the placement of the pole-part operation.  It is
tempting to treat the last term on the first line of this equation as
$(C_{0} ZK_{0}) \polepart$.  But this would mean that the pole-part
operation would be applied to the whole object $C_{0} ZK_{0}$, whereas
it should only be applied to the quantity that is an operator matrix
element, i.e., to $ZK_{0}$; this is indicated by the brackets.  The
incorrect method, of taking the pole part of everything, i.e., of
$C_{0} ZK_{0}$, will get different results from the correct method if
$C_{0}$ has any dependence on the regulator parameter $\epsilon
$---see the Appendix.

For the general case, we apply the factorization theorem to a target
which is a single on-shell parton.  The structure function in this
case, $F_{p}$, is obtained by setting $D=0$ and $T_{0}=Z$ in Eq.\
(\ref{2PI}), and it follows that the remainder term $r$ is zero ---
see Eq.\ (\ref{remainder}).  We let $A_{Bp}$ and $A_{Rp}$ correspond
to parton densities on a parton target:\footnote{
   Observe that the word ``parton'' has just been used with two
   different meanings.  The parton target is an on-shell state
   corresponding to one of the elementary fields in the Lagrangian.  A
   parton density is a number density computed using a particular
   operator involving the corresponding field.  Thus a parton density
   in a parton is a non-trivial but non-contradictory concept. 
}
\begin{equation}
   A_{Bp} = Z \frac {1}{1-K_{0}} Z , ~~~ A_{Rp} = Z \frac {
1}{1-K_{0}(1-\polepart)} Z .
\end{equation}
Then the bare factorization theorem
Eq.\ (\ref{factorization.bare}) becomes just\footnote{
   Note that this equation has no remainder term even if we have
   non-zero quark masses, since we have not yet taken a zero-mass
   limit in the coefficient function.  To compute the coefficient
   function for a light parton, it is normally convenient to take
   the zero mass limit, as we will see later.  In that case the
   remainder term on a parton target will become nonzero.
}
\begin{equation}
   F_{p} = C_{B} \otimes A_{Bp} ,
\label{bare.factorization.parton}
\end{equation}
while the renormalized factorization theorem on a parton target is
\begin{equation}
   F_{p} = C_{R} \otimes A_{Rp} .
\label{factorization.parton}
\end{equation}
Neither of these equations has a remainder term.  The coefficient
function is, of course, target independent; it is the same here, on a
parton target, as in the factorization theorem on a hadron target.

We expand in powers of $K_{0}$, and the $n$th term in $F_{p}$ is
\begin{equation}
   F_{p}^{(n)} = C_{R}^{(n)} + \sum _{j=0}^{n-1} C_{R}^{(j)} A_{Rp}^{(n-j)} .
\label{coefficient.recursion.0}
\end{equation}
Rewriting this equation as
\begin{equation}
    C_{R}^{(n)} = F_{p}^{(n)} - \sum _{j=0}^{n-1} C_{R}^{(j)} A_{Rp}^{(n-j)} .
\label{coefficient.recursion}
\end{equation}
gives the desired recursion.  The $n$th order renormalized coefficient
is the $n$th order partonic structure function minus lower-order terms
in the Wilson coefficients times partonic matrix elements of the
operators defining the parton densities.  Both the partonic structure
functions and the partonic operator matrix elements can be computed in
perturbation theory, and actual calculations to order $\alpha _{s}^{2}$
exist \cite{BMSvN}.  The recursion starts at order 0, where the
coefficient function is the lowest-order partonic structure
function: the first non-trivial case, for $n=1$, is exactly Eq.\
(\ref{first.order.coeff}).

The indices $n$ and $j$ can equally well be interpreted as
parameterizing an expansion in loops (or $\alpha _{s}$) as well as an
expansion in powers of $K_{0}$.

\section{Parton densities}
\label{sec:pdf}

\subsection{Gauge-invariant parton-densities}

Our derivation leads to a factorization theorem in which the
bare parton densities are defined by formulas like
\begin{equation}
   f_{B}(x) = \int  \frac {dy^{-}}{2\pi } e^{-ixp^{+}y^{-}}
          \langle p| \overline\psi (0,y^{-},{\bf 0}_{T})
          \gamma ^{+}  \psi (0) |p\rangle .
\label{pdf.bare.lc}
\end{equation}
(The vacuum expectation value of the operator should be
subtracted, so that this matrix element is a connected one.)
In a gauge theory like QCD, this is a matrix element of a
gauge-variant operator.  The gauge to be used to define the
operator is the light-cone gauge $A^{+}=0$, since that was the gauge
used for the proof of factorization.  In accordance with the
derivation, the two quark fields are {\em renormalized} quark
fields.  However, as we saw, there are divergences associated
with the bilocal light-cone operator, so this formula, without
renormalization, defines a bare parton density.\footnote{
   A better definition of a bare parton density is to replace the
   renormalized quark fields by bare quark fields.  This new
   definition differs from the one given above by a factor of the
   quark's wave-function renormalization.  The advantage of this second
   definition is that it is renormalization-group invariant, so that
   formal derivations of the renormalization-group equation are simpler.
}

As is well known,
a gauge invariant form of the parton density can easily be
made by inserting a path-ordered exponential of the gluon field:
\begin{equation}
   f_{B}(x) = \int  \frac {dy^{-}}{2\pi } e^{-ixp^{+}y^{-}}
          \left\langle  p \left| ~\overline\psi (0,y^{-},{\bf 0}_{T})
            P {\rm exp}
            \left[
                -ig_{0} \int _{0}^{y^{-}} d{y'}^{-}
                 t_{a}A_{0a}^{+}(0,{y'}^{-},{\bf 0}_{T})
            \right]
          \gamma ^{+} \psi (0)~  \right| p \right\rangle .
\label{pdf.bare.gi}
\end{equation}
In the light-cone gauge $A^{+}=0$, the exponential reduces to unity,
so that the parton density agrees with the previous definition. Note
that to get gauge invariance the coupling and the gluon field in the
exponential are the bare ones.

Renormalization is performed by convoluting the bare parton
densities with the previously determined renormalization factor.

Notice that the recursion formula, Eq.\
(\ref{coefficient.recursion}), for the coefficient function is
actually gauge invariant, if we interpret it as an equation for
terms in expansions in powers of $\alpha _{s}$.  For example, the
left-hand side is the $\alpha _{s}^{n}$ term in the expansion of the
structure function of an on-shell quark or gluon, and the
coefficients $A_{Rp}^{(n-j)}$ are terms in the expansion of the
renormalized parton densities in the same on-shell quark or gluon
state.

\subsection{Evolution equations}

The final element in the factorization formalism that makes it
useful for phenomenology is the set of DGLAP evolution equations.
Since the parton densities are matrix elements of renormalized
composite operators, the evolution equations are just the
ordinary renormalization-group equations for the operators.
To use the factorization formula one sets the
renormalization/factorization scale $\mu $ to be of order $Q$.  Then
there are no large logarithms in the coefficient functions, for
which low-order perturbation calculations are therefore useful.
The parton densities at different scales are related by use of
their evolution equations.

Since we have chosen to use \MSbar{} renormalization, the
renormalization-group coefficients are independent of masses, and
are in fact the ones normally used.  This is true even if one (or
more) of the quarks is heavy and has a mass $M$ comparable with
$Q$.  Our proof of factorization has demonstrated that all
relevant effects of non-zero quark masses can be found either in
the coefficient functions or in the starting values of the parton
densities.

Of course, one can perturbatively compute the values of the heavy
quark densities, by the methods that Witten \cite{Witten} first
devised.  In our formalism this is most conveniently done in
association with the version of factorization that is appropriate
when $M$ is bigger than $Q$, which we will treat in Sect.\
\ref{sec:small.Q}.

\section{Quark masses in the coefficient function}
\label{sec:masses}

In conventional treatments of factorization, masses are set to
zero in the coefficient functions.  But our treatment has
preserved masses, and this is the key to a correct treatment of
the effects of heavy quarks.

\subsection{Massless limit}

The massless limit can be taken in the coefficient function.  This can
be done since the $1-Z$ factors in Eq.\ (\ref{coefficient.ren}) cancel
leading power contributions from all regions except where all the loop
momenta are of order $Q^{2}$ in virtuality, and except for regions
that contribute to the (canceled) UV divergences.  Thus setting a
mass $m$ to zero gives an error that is a power of $m/Q$. A particular
consequence of this result is that all potential collinear divergences
are canceled. Thus the coefficient function is a truly infra-red safe
quantity. If the renormalization mass $\mu $ is chosen to be of order
$Q$, then perturbative calculations can be made.

Since errors in setting a mass to zero are a power of $m/Q$,
taking the massless limit is sensible if all the quark masses are
of the order of a typical hadronic mass or smaller; the errors
are no bigger than errors that have been made elsewhere in the
derivation of factorization.

\subsection{Heavy quarks}

However, there are quarks whose masses are larger than this (charm,
etc.).  Let us first treat the case that there is only one heavy quark,
of mass $M$. It is not always appropriate to set $M=0$ in the
coefficient functions, since the error in doing so is of order
$(M/Q)^{p}$, which may be much bigger than the error associated with
dropping the remainder term in the derivation of the factorization
theorem.  A error of order $(M/Q)^{p}$ may also be larger than the
error caused by using a finite order truncation of the perturbation
series for the coefficient function.

Now, the error in the factorized form of the structure function
is of order $(\Lambda /Q)^{p}$, and the derivation
of this error estimate is valid over the whole range of quark
mass for which $Q \gtrsim M$. This means both the region where
$Q$ is of order $M$ and the region where $Q$ is much bigger than
$M$.  The remainder term is uniformly suppressed by a power of
$\Lambda /Q$.  The sole effect of a heavy quark line is to restrict its
virtuality to be at least of order $M^{2}$, and this is completely
compatible with the derivation of the error estimate.

We therefore have a factorization theorem that is valid in the whole
of the region that $Q \gtrsim M$, as we have already observed.  If $Q$
is sufficiently much bigger than all the quark masses, then we may set
all the masses to zero in the coefficient function.  If some of the
quark masses are non-negligible, then we simply leave their masses at
their correct values.

However, these considerations only apply if $M \lesssim Q$.  If, on
the contrary, a heavy quark mass is much larger
than $Q$, then the coefficient functions that we constructed
have logarithms of $M/Q$ in this region of relatively small $Q$.
This is a problem we will treat in Sec.\ \ref{sec:small.Q}.  The
work in this section is based on a factorization theorem derived
under the condition that $Q$ is at least comparable with $M$.

Despite the fact that we have retained heavy quark masses wherever
necessary, the kernels of the the evolution equations for the parton
densities are in fact the same as with the quark masses
set to zero, i.e., they are identical to the ordinary DGLAP
equations in the \MSbar{} scheme.  This happens because the
evolution equations are in our approach just the renormalization
group equations for the renormalized parton densities.  The
Altarelli-Parisi kernels are anomalous dimensions, obtained from
the renormalization factor $G_{ij}$.  Since the renormalization
counterterms in the \MSbar{} scheme are mass independent, so are
the Altarelli-Parisi kernels, a statement that is true not only
for the leading-order $\alpha _{s}$ terms in the kernels, but for all
higher order corrections.

\subsection{Redefinition of the $Z$ operation}

The analytic core of our proof is in the definition of the $Z$
operation and in the proof that the remainder term, Eq.\
(\ref{remainder}), is suppressed by a power of $\Lambda /Q$.  The rest of
the proof is simple linear algebra.  It is possible to adjust to
the definition of $Z$ to make calculations more convenient. We
have already made one such redefinition---see Eqs.\
(\ref{Z.def1}) and (\ref{Z.def.massive}).

In the next section we will propose one further redefinition of
$Z$ that will simplify some calculations, by allowing heavy
quark masses to be set to zero in certain parts of the
calculations of the coefficient functions.
But first we must characterize the allowed redefinitions.  We
address explicitly only the momentum dependence of $Z$.  The
spin-dependent part can be discussed in a similar fashion.

The first and most essential property is that $Z$ provide a good
approximation to leading regions, of the form of Fig.\
\ref{fig:leading.region}, i.e., that
\begin{equation}
   H\cdot Z\cdot T = H \cdot  T + \mbox{non-leading power} ,
\end{equation}
whenever we are in an integration region where
the virtualities in $H$ are much bigger than the
virtualities in $T$.  The second property is that when we go
outside the momenta for which $Z$ gives a good approximation,
insertion of a factor of $Z$ should not produce a result that is
much bigger than the original.  To make this precise, let $H$ and
$T$ be subdiagrams that could be used in Fig.\
\ref{fig:leading.region}.  We have
\begin{equation}
   H \cdot  T = \int  \frac {d^{4}k}{(2\pi )^{4}} H(q,k) T(k,p)
\end{equation}
and
\begin{equation}
    H \cdot  Z \cdot  T = \int  \frac {d^{4}k}{(2\pi )^{4}} \int  \frac {
d^{4}l}{(2\pi )^{4}}
                H(q,k) Z(k,l) T(k,p) .
\end{equation}
We require, with one exception, that $H\cdot Z\cdot T$ should not be much
larger than $H\cdot T$.  The exception is that we can have a
logarithmic ultra-violet divergence for large $l^{2}$.

The above properties are sufficient to ensure that the remainder
as defined in Eq.\ (\ref{remainder}) is power suppressed.  Then
we can obtain the renormalized factorization theorem Eq.\
(\ref{factorization.ren}) given that any divergences in the
operator matrix elements are at worst logarithmic.

A final property is needed in order that the factorization theorem be
of a usefully simple form.  We choose this to mean that factorization
involves a convolution in just one variable, a longitudinal momentum
fraction. This forces the momentum-dependent part $Z$ to be of the
form
\begin{equation}
   Z(k,l) = \delta ^{(4)}\left[k^{\mu } - \hat l^{\mu }\right] f(l) .
\end{equation}
Here the function $f(l)$ must be unity when $l_{T}$ is less than about
$Q$ and $l^{-}$ is less than about $Q^{2}/p^{+}$.  Moreover, the
approximated momentum $\hat l^{\mu }$ must approach $(l^{+},0,{\bf
0}_{T})$ in the collinear limit.  Both $f(l)$ and $\hat l^{\mu }$ must
be smooth functions.  In order that the convolution in the
factorization formula be a convolution in one variable, the
approximated momentum $\hat l^\mu$ must be independent of $l^-$ and
$l_\perp$.  

Perhaps the simplest and most natural definition is to write
\begin{equation}
   Z(k,l) = \delta ^{(4)}\left[k^{\mu } - (k^{+}, 0, {\bf 0}_{T})\right]
            \theta (l_{T}<\mu ) ,
\end{equation}
which is just like Eq.\ (\ref{Z.def1}), except for a cut-off on
the transverse momentum entering from the right.  This definition
would be favored, for example, by Brodsky \cite{Brodsky}.  It
corresponds to defining parton densities by integrals of the
following form:
\begin{equation}
   f(x,\mu ) = \mbox{standard normalization factors} \times
     \int _{-\infty }^{\infty } dl^{-} \int _{l_{T}<\mu } d^{2}{\bf l}_{T}
    ~\left\langle p \left|
      \, \overline\psi (-l) \, \gamma ^{+} \,\psi (l)
    \, \right| p \right\rangle ,
\end{equation}
where there is an integral over all virtualities of the parton
from the target and an integral up to a certain maximum
transverse momentum, and we are using the Fourier-transformed
fields.

This definition suffers from two inconveniences.  The first is that in
a gauge theory it does not give parton densities that are manifestly
gauge invariant.  The second is that the evolution equations (in $\mu
$) are not exactly homogeneous equations of the Altarelli-Parisi form;
a subsidiary expansion for large $\mu $ is needed to get the
Altarelli-Parisi equations.

Neither disadvantage is fatal, but we prefer to use a definition in
which $f(l)=1$, as in Eqs.\ (\ref{Z.def1}) and (\ref{Z.def.massive}).
The parton densities are then precisely of the form of light-cone
operators, and UV renormalization must be applied as described in
earlier sections.

\subsection{Proposal for optimal redefinition of $Z$}
\label{sec:redefinition.of.Z}

The remaining freedom in defining $Z$ resides in what it does to the
factors on its left, and in the definition of the approximated
momentum $\hat l$.  The most natural definition is perhaps the one in
Eq.\ (\ref{Z.def.massive}).  But a simplification is possible.

Let us first recall the classification of partons as light or
heavy according to whether their masses are less than or greater
than a few hundred MeV.  Thus the gluon, and the up, down, and
strange quarks are light, while the charm, bottom, and top quarks
are heavy.  The importance of this distinction is that it is
always legitimate to neglect light parton masses in the hard
scattering coefficients, since the errors in doing so are of the
same order as the non-leading power corrections (``higher-twist
terms'') that constitute the remainder in the factorization
formula.  But it is not always valid to neglect heavy quark
masses.  Even if $Q$ is much larger than the mass $M$ of some
heavy quark, the error resulting from replacing $M$ by zero in
the coefficient function is larger than the errors that result
from neglect of higher twist terms.  (In practice we normally
have larger errors that result from truncation of the
perturbation expansion of the coefficient functions, and then it
will be sensible to neglect $M$ at suitably high $Q$.)  Note,
however, that it is never legitimate to neglect masses in the
parton density.

So it is convenient to equip $Z$ with a prescription to set light
parton masses in everything to its left.  This new operation we call
$Z_{1}$.  Consider a convolution $H\cdot T$ like that implied by Fig.\
\ref{fig:leading.region}, and suppose that $H$ and $T$ are joined by a
pair of light parton lines.  We have
\begin{equation}
   H\cdot T = \int d^{4}k \, H(q,k,m,M) \, T(k,p,m,M) ,
\end{equation}
so that
\begin{equation}
   H \cdot  Z_{1} \cdot  T = \int d\xi  \, H(q,\xi \hat p, 0,M)
                \int d^{2}{\bf k}_{T} \, dk^{-} \, T(k,p,m,M),
\label{Z1.def.light}
\end{equation}
where $\hat p = (p^{+},0,{\bf 0}_{T})$, and, for simplicity, we have
omitted the treatment of the Dirac matrices, which is unchanged
from our earlier work. We use $m$ and $M$ to refer to light
and heavy parton masses.

In the above equations, we have assumed that the limit of zero
mass for the light partons exists.  This is, of course, normally
not true if $H$ is a simple sum of Feynman graph, such as
corresponds to the $H$ subgraph in Fig.\
\ref{fig:leading.region}.  Rather $H$ should be a quantity such as
a bare coefficient function obtained from such a subgraph with a
series of subtractions to cancel the collinear regions, i.e., a
quantity such as
\begin{equation}
   C_{0} \cdot  \frac {1}{1 - (1-Z_{1})K_{0}} .
\label{subtracted.object}
\end{equation}

Just like the pole part operation, $\polepart$, $Z_{1}$ is not a
linear operator, at least not on momentum space.  Nevertheless it
obeys enough of the algebraic rules for linear operators that the
proof of factorization still works if we replace $Z$ by $Z_{1}$.
The advantage of the use of $Z_{1}$ is that it directly implements
the zero-mass limit for light partons in the definition of the
coefficient functions.  It is necessary to add to the proof a
verification that the zero-mass limit is only being applied to
quantities for which the limit exists, at all stages of the
proof.  The verification is elementary, since the dangerous
regions arise from regions of exactly the kind that are
suppressed by the $1-Z_{1}$ factors  in Eq.\
(\ref{subtracted.object}).  We can apply the same arguments to the
renormalized coefficient functions as well.

In practical work, it is of course very important to take the
zero mass limit wherever possible, since massless Feynman graphs
are generally much easier to calculate than massive ones.

We now show that there are certain parts of calculations with heavy
quarks where one can correctly redefine $Z_{1}$ also to set {\em
heavy} quark masses to zero, even when $Q$ is of order $M$.  Let us
continue to define $Z_{1}$ as in Eq.\ (\ref{Z1.def.light}) when the
lines joining $H$ and $T$ are light partons.  The light parton masses
are set to zero in $H$, but the heavy parton masses are not.

But now suppose $H$ and $T$ are joined by heavy quarks.  We will
now show that it is legitimate to define $Z_{1}$ to set the heavy
quark mass(es) to zero in $H$:
\begin{equation}
   H_{Q} \cdot  Z_{1} \cdot  T_{Q} = \int d\xi
                ~ H_{Q}(q,\xi \hat p, 0,0)
                \int d^{2}{\bf k}_{T} \, dk^{-} \, T_{Q}(k,p,m,M).
\label{Z1.def.heavy}
\end{equation}
Here we have equipped $H$ and $T$ with a subscript $Q$ to
symbolize their being joined by heavy quark lines.

\begin{figure}
    \begin{tabular}{ccc}
           \epsfxsize=1.5in
           \epsfbox{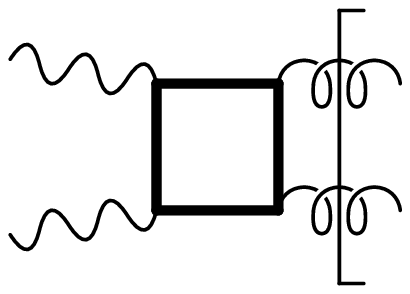}
        &
           \epsfxsize=2.5in
           \epsfbox{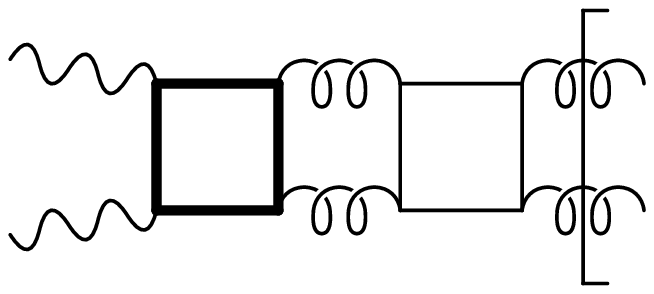}
        &
           \epsfxsize=2.0in
           \epsfbox{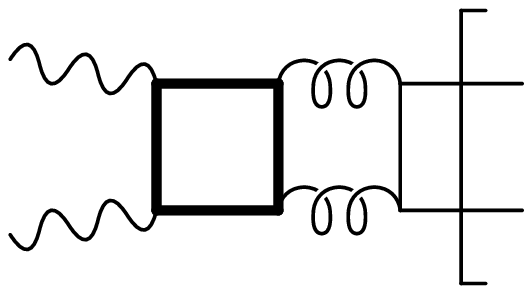}
    \\
      (a) & (b) & (c)
    \\*[0.2in]
           \epsfxsize=1in
           \epsfbox{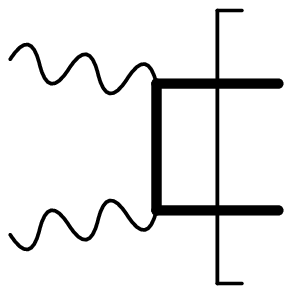}
        &
           \epsfxsize=1.5in
           \epsfbox{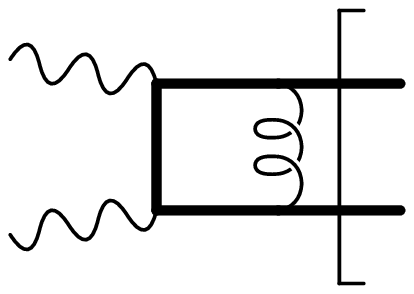}
        &
           \epsfxsize=2.0in
           \epsfbox{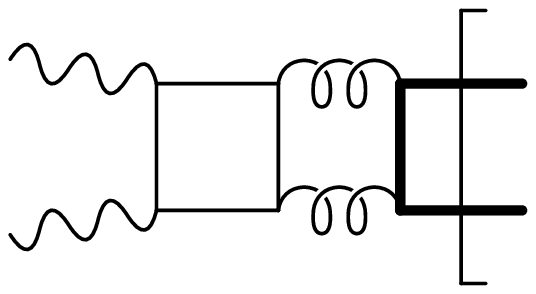}
    \\
      (d) & (e) & (f)
    \end{tabular}
    \vspace*{0.2in}
\caption{Diagrams with the $Z_{1}$ operation applied at the vertical
        line. The heavy quarks are denoted by the thick solid
        lines.}
\label{fig:Z1.examples}
\end{figure}

In Fig.\
\ref{fig:Z1.examples} we show some diagrams to which $Z_{1}$ is
applied, at the place indicated by the vertical line.  To allow
zero mass limits to be taken, we assume implicit $1-Z_{1}$ factors
at all necessary points to the {\em left} of the vertical bar,
as in Eq.\ (\ref{subtracted.object}).
In the case that there is more than one heavy quark, one should
set to zero only the masses of those quarks that are lighter than
the quarks joining $H$ and $T$.  This need for this last
requirement will become apparent in the proof.

In the first three graphs, which have either gluons or light
partons as their external lines, only the light quarks have their
masses set to zero.  But in the last three graphs, which have
heavy quark external lines, all the quark masses should be set to
zero; the external quarks will also be given massless on-shell
momenta, $k^{2}=0$, and the Dirac matrix will be that for a massless
quark.

If it is indeed valid to define $Z_{1}$ in this way, a substantial
simplification is achieved in practical calculations, since it is
only necessary to retain non-zero masses for heavy quarks in
loops of heavy quark lines in coefficient functions with external
light lines, i.e., in graphs such as the first three of Fig.\
\ref{fig:Z1.examples}.

The formal proof is as follows.
\begin{enumerate}

\item
    $H\cdot Z_{1}$ is only used when $H$ has a zero mass limit.  Hence
    the virtualities in $H$ are of order $Q^{2}$ or larger.  This is
    simply the assertion that collinear subtractions have been
    applied inside $H$, as in Eq.\ (\ref{subtracted.object}).

\item
    If $H$ and $T$ are joined by heavy quark lines, the
    virtuality of the heavy quark is at least of order $M^{2}$ in
    the dominant region of integration, for the whole leading
    power.  The virtuality, as is well known, is in fact
    space-like.

\item
    In a region where the virtualities in $T$ are much less than
    the virtualities in $H$, then $H\cdot Z_{1}\cdot T$ provides as good an
    approximation to $H\cdot T$ as does the approximation with the
    heavy quark mass left non-zero.  The original approximation
    involved replacing a momentum of space-like virtuality of
    order $M^{2}$ by an on-shell momentum.  Instead we now replace
    it by a light-like momentum.  The new $Z$ operation provides
    a suitable approximation given that the old operation did.
    Thus the first essential property of a $Z$ operation is
    obeyed

\item
    If the virtuality of the lines joining $H$ and $T$ is of
    order the virtualities in $H$, then setting masses to zero in
    $H$ changes the precise value but not the order of magnitude.
    Thus $H\cdot Z_{1}\cdot T$ is of the same magnitude as $H\cdot T$ in this
case.
    The second property for $Z$ is satisfied.

\item
    The effect of $Z_{1}$ on $T$, in $Z_{1}\cdot T$, is the same as for $Z$.
    Thus there is no change in the logarithmic UV divergences
    that are generated.

\end{enumerate}

\begin{figure}
   \begin{displaymath}
      \epscenterbox{1.5in}{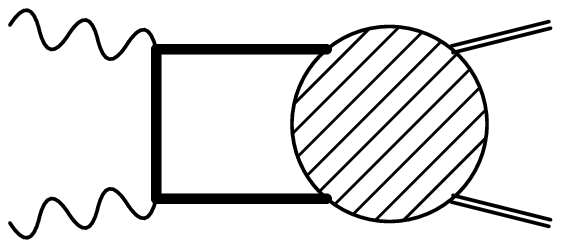}
      ~~~\longrightarrow~~~
      \epscenterbox{1.5in}{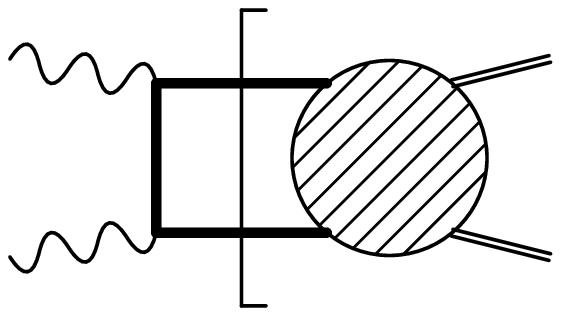}
   \end{displaymath}
\caption{Born graph for heavy quark in DIS.}
\label{fig:Q.0}
\end{figure}

A more physical argument can be made with the aid of an example.
Consider the lowest order calculation of a heavy quark loop to a
structure function.  In Fig.\ \ref{fig:Q.0}, we have the Born graph
for DIS on a heavy quark that comes out of the shaded target bubble.
If $Q$ is much larger than the quark mass $M$, it is a useful
approximation to replace the graph by the lowest order Wilson
coefficient times the heavy quark density, as shown on the right of
Fig.\ \ref{fig:Q.0}, for the important region when the quark has
transverse momentum much less than $Q$. It is also a good
approximation to replace $M$ by zero in the contribution to the
coefficient function.

\begin{figure}
   \begin{eqnarray}
      \epscenterbox{1.8in}{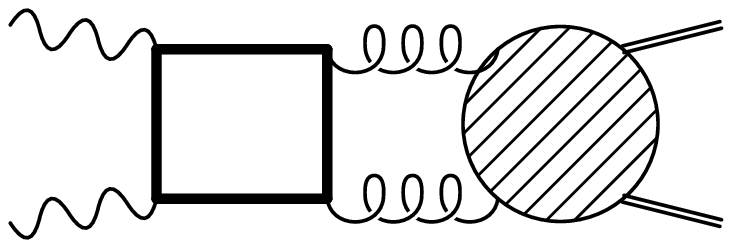}
      &~~~\longrightarrow~~~&
      ~\left[ ~
          \epscenterbox{1.1in}{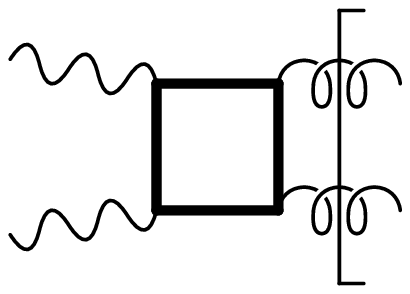}
          ~~-~~ \epscenterbox{1.1in}{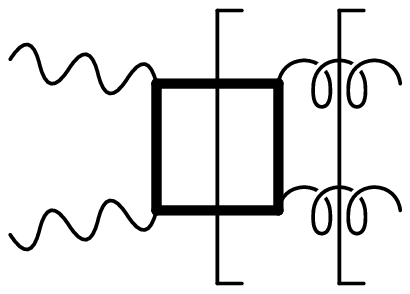}
      ~ \right]
      \epscenterbox{0.9in}{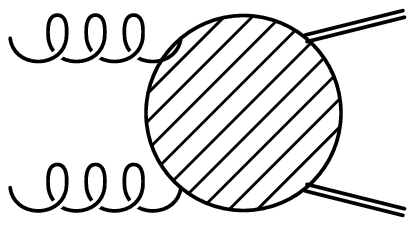}
   \nonumber\\
      && + ~~\epscenterbox{1.8in}{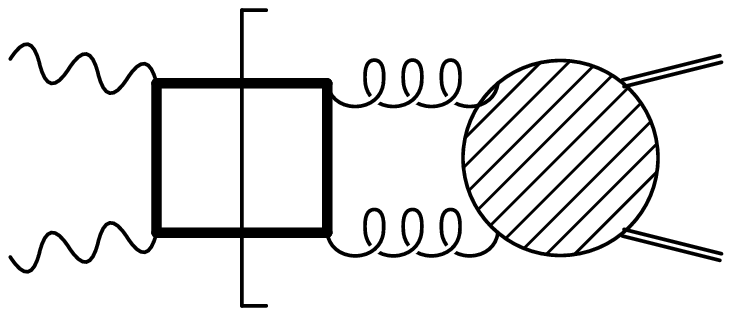}
   \nonumber
   \end{eqnarray}
\caption{One-loop graph for heavy quark in DIS.}
\label{fig:Q.1}
\end{figure}

Now both of these approximations fail when $Q$ is comparable to $M$
(the ``threshold region'').  But in this case the heavy quark
distribution is of order $\alpha _{s}$ relative to the gluon
distribution.  So to do valid phenomenology we must include also a
one-loop coefficient times the gluon density.  The result is shown in
Fig.\ \ref{fig:Q.1}.  We start with a particular kind of graph for the
structure function where a heavy-quark loop couples to the target by
gluon lines. To avoid extra irrelevant complications, suppose that the
gluons have low virtuality. The first part of the right-hand side is a
contribution to the coefficient function times the gluon density.  In
the one-loop coefficient there is a subtraction term.  The second term
on the right is the previously defined heavy quark coefficient
function times a heavy quark density.  In the region where the gluons
have low virtuality this second term cancels the subtraction in the
one-loop Wilson coefficient times the gluon density.

Hence the incorrectness in the approximation used in Fig.\
\ref{fig:Q.0} is compensated by the subtraction in Fig.\
\ref{fig:Q.1}.
Of course, it would have been much simpler to use the heavy quark
(or ``fixed-flavor'') scheme that we will discuss in Sec.\
\ref{sec:small.Q}.  But that scheme does not permit us to go to large
$Q$, because there will then be large logarithms of $Q/M$ in its
coefficient functions.  In contrast, the scheme in Figs.\
\ref{fig:Q.0} and \ref{fig:Q.1} permits an interpolation between
low and high $Q$ {\em without loss of accuracy}.

At sufficiently large $Q$, the Born term alone provides useful
phenomenology, because the heavy quark density is large.
Moreover, zero-mass coefficient functions can be used.

As $Q$ is decreased towards the threshold region, the Born term
in the coefficient function becomes increasingly inaccurate as a
representation of the graph on the left of Fig.\ \ref{fig:Q.0}.
Note that even if we evaluate the coefficient with the correct
mass there is still an error of the same order of magnitude as
the error in neglecting the mass completely.  This is because the
horizontal lines are necessarily space-like.  Replacing them with
on-shell lines gives an error of order $M^{2}/Q^{2}$.

When one decreases $Q$, the errors in the approximation of Fig.\
\ref{fig:Q.0} increase.  At order $\alpha _{s}$, the errors are
compensated by the subtraction term in Fig.\ \ref{fig:Q.1}. But
beyond some point, the errors in the approximation become larger
than the quantity one is trying to compute.  Correct compensation
of errors will involve the use of even higher order diagrams.
Then one must abandon this scheme and use only the heavy quark
scheme of Sec.\ \ref{sec:small.Q}.  The important point is that
there is an overlap in the region of validity of both schemes.

\section{Factorization with $Q \protect\lesssim M$}
\label{sec:small.Q}

When $Q$ is reduced below the mass $M$ of a heavy quark, the scheme
described in Sec.\ \ref{sec:large.Q.proof} becomes inappropriate.
Indeed, given a fixed value of $x$, we go below the threshold at
$Q=2M\sqrt{x/(1-x)}$ for producing the heavy quark by reducing $Q$
enough.  On the other hand the factorization theorem that we derived
earlier has a non-zero subprocess in which there is production of
heavy quarks in the final state, for any value of $Q$.  An example is
given by Fig.\ \ref{fig:Q.0}.  There we replace a graph for heavy
quark production by the lowest order approximation to the
factorization theorem.  The replacement of an off-shell heavy quark by
an on-shell quark in the hard scattering enables the approximated
graph to be non-zero, even when the true physical process is below the
threshold for producing heavy quarks.  The error in the approximation
is repaired by higher-order approximations to the coefficient
functions, as illustrated in Fig.\ \ref{fig:Q.1}.

Clearly it is likely to be a poor and inaccurate method of calculation
to obtain an answer that is known to be zero by adding a collection of
non-zero pieces, in a truncated perturbation expansion. Even a little
above threshold we may have inaccurate calculations: a cross section
that approaches zero as the threshold is approached is calculated as a
sum over terms that do not have the correct threshold behavior.

The remedy is to use a different version of the factorization theorem,
in fact the well-known fixed-flavor-number scheme \cite{F2c,DEN}.  In
this section we present a proof of factorization in this scheme in a
form that will mesh with the formulation and proof of factorization
that we gave earlier. Using the terminology introduced in Sec.\
\ref{sec:CWZ}, we will say that the heavy quark is treated as
non-partonic. It will be convenient, for the purposes of this section,
to call this scheme the ``heavy-quark scheme''.  The essence will be to
treat the heavy quark as always being part of the hard
scattering. This scheme has a range of validity that includes the
whole region that $Q \lesssim M$.  This range overlaps with the range
of validity of the factorization theorem where the heavy quark is
treated as partonic, i.e., the range $Q \gtrsim M$.

There are two important observations.  One is that when $Q$ is of
order $M$, the heavy quark mass provides a large scale of virtuality
that can be treated on the same footing as $Q$.  The second
observation is that when $Q$ is much less than $M$, the decoupling
theorem \cite{AC} applies.  Our heavy quark scheme will satisfy the
decoupling theorem in the simplest way: one can simply drop all graphs
involving heavy quark lines and obtain a correct answer without needed
extra finite renormalizations of the coupling and parton densities.
The method we will use is that of Collins, Wilczek and Zee \cite{CWZ},
with the heavy quark begin treated as non-partonic. In that subscheme,
renormalization is done in the \MSbar{} scheme for all graphs except
those involving the heavy quark. For graphs with a heavy quark loop,
renormalization is done by subtraction at zero momentum and with the
light quark masses set to zero. The remaining renormalizations involve
graphs with external heavy quark lines. Following Buza {\it et al.}\
\cite{BMSvN}, we define the heavy quark mass as the position of the
pole in the heavy quark propagator, a definition that makes sense in
perturbation theory. Remaining renormalizations are defined by
pole-part subtractions, in the \MSbar{} style.

This advantages of this scheme are \cite{CWZ}:
\begin{enumerate}

\item
    It satisfies manifest decoupling.

\item
    \MSbar{} and zero momentum subtraction allow preservation of
    Ward identities in gauge theories without the need for extra
    finite counterterms.

\item
    Anomalous dimensions for the active partons and the $\beta $
    function are the same as in the \MSbar{} scheme for the
    theory with the heavy quark omitted.  They have no mass
    dependence.

\item
    At no stage, in either this subscheme or the subscheme where the
    heavy quark is active (or partonic), do we have to make an expansion
    in powers of $M/Q$ or $Q/M$: the heavy quark mass need never be
    approximated. So the scheme can be applied when there are several
    heavy quarks and the ratios of their masses are not necessarily
    large. Furthermore, there is no loss of accuracy when treating
    problems where a heavy quark is not heavy enough for it to decouple
    to high accuracy and not light enough for its mass to be
    approximated by zero. 

\end{enumerate}

In this section we will treat the case that the theory contains
one heavy quark and that $Q \gtrsim M$.  The most general case,
that there are several heavy quarks, whose masses may or may not
be larger than $Q$, will form an elementary generalization to be
treated in Sec.\ \ref{sec:multiple}.
We will first derive a factorization theorem without taking
account of renormalization and then we will do the
renormalization.

\subsection{Bare factorization theorem}

When we are in the region $Q \lesssim M$, the leading regions continue
to be of the form of Fig.\ \ref{fig:leading.region}.  However, the
specification of the graphs is a bit different, since heavy quark
loops must each be contained in the hard part $H$ or in
renormalization subgraphs of $T$.  Thus the lines joining $H$ and $T$
must always be light partons. To obtain a factorization theorem, we
use the reasoning in Sec.\ \ref{sec:large.Q.proof} with two changes.

The first change is that since heavy quarks cannot join the hard and
target subgraphs, we change Eq.\ (\ref{2PI}) so that the amplitudes
corresponding to $C_{0}$, $K_{0}$, $T_{0}$ and $D$ are two-particle
irreducible in the light partons only. The second change is that we
need to take account of the decoupling theorem for graphs with heavy
quark loops.

The first change means that Eq.\ (\ref{2PI}) needs to be replaced by
\begin{eqnarray}
    F &=& \sum _{n=0}^{\infty } C_{H} \cdot  (K_{H})^{n} \cdot  T_{H} + D_{H}
\nonumber\\
      &=& C_{H} \cdot  \frac {1}{1-K_{H}} \cdot  T_{H} + D_{H} ,
\label{2PI.light}
\end{eqnarray}
where the subscript $H$ means that the amplitude with the subscript is
2PI only in light parton lines.  We can formalize the definitions of
$C_{H}$, etc.\ by defining a projection $P_{L}$ that is unity on light
lines and zero on heavy quark lines.  The projector onto heavy lines
is $P_{H} = 1-P_{L}$.  Then
\begin{eqnarray}
   C_{H} &=& C_{0} \cdot \frac {1}{1-P_{H} K_{0}} \cdot P_{L} ,
\nonumber\\
   K_{H} &=& P_{L} \cdot K_{0} \cdot \frac {1}{1-P_{H} K_{0}} \cdot P_{L}
        = P_{L} \cdot \frac {1}{1-K_{0} P_{H}} \cdot K_{0} \cdot P_{L} ,
\nonumber\\
   T_{H} &=& P_{L} \cdot \frac {1}{1-K_{0} P_{H}} \cdot T_{0} ,
\nonumber\\
   D_{H} &=& D_{0} 
   + C_{0} \cdot \frac {1}{1-P_{H} K_{0}} \cdot P_{H} \cdot
T_{0} .
\end{eqnarray}
It can be verified that with these definitions, the structure function
given by Eq.\ (\ref{2PI.light}) is the same as before, i.e.,
$C_{0} \cdot  \frac {1}{1-K_{0}} \cdot  T_{0} + D_{0}$.

We define the remainder to be
\begin{equation}
 r_{H} = C_{H} \cdot  \frac {1}{1 - (1-Z)K_{H}} \cdot  (1-Z) \cdot  T_{H} +
D_{H}.
\label{remainder.H}
\end{equation}
This remainder is power suppressed, just like the remainder $r$ that we
defined in Eq.\ (\ref{remainder}).

No changes are needed in the reasoning that lead to the bare
factorization theorem Eq.\ (\ref{factorization.bare}).  We find
that
\begin{equation}
   F = C_{HB} \otimes A_{B} + \mbox{non-leading power} ,
\label{factorization.bare.H}
\end{equation}
where the bare coefficient function is
\begin{eqnarray}
   C_{HB} &=& C_{H} \cdot  \frac {1}{1 - (1-Z)K_{H}} \cdot  Z ,
\nonumber\\
       &=& C_{H} \cdot  \frac {1}{1 - (1-Z)K_{H}}
              \cdot  Z \cdot  P_{L},
\label{coefficient.bare.H}
\end{eqnarray}
and the bare operator matrix element (or bare parton density) is
\begin{eqnarray}
   P_{L} \cdot  A_{HB} &=& P_{L} \cdot  Z
          \cdot  \frac {1}{1-K_{H}} \cdot  T_{H} .
\label{operator.bare.H}
\end{eqnarray}

The leading regions only have active, light partons joining the hard
subgraph and the target subgraph.  This is reflected in the formulas by
the fact that there are explicit factors of $P_{L}$ on the right of $C_{H}$,
on the left of $T_{H}$ and on both sides of $K_{H}$. Hence we may insert
the explicit factors of $P_{L}$ in the formulas for the coefficient
function and operator matrix elements, Eqs.\ (\ref{coefficient.bare.H})
and (\ref{operator.bare.H}).

The reader should clearly understand the distinction between the
following notations: $C_{0}$ is a fully 2PI and amputated Green function
for two virtual photons and two quarks; $C_{H}$ is the same Green
function as $C_{0}$ except for being 2PI only in light parton lines; and
finally $C_{HB}$ is a Wilson coefficient: it is the full amputated
Green function, including reducible graphs but with subtractions to
make it a purely UV object.

Contrary to appearances, the definition Eq.\ (\ref{operator.bare.H})
is equivalent to the previous definition, Eq.\
(\ref{operator.bare}), so that
\begin{eqnarray}
   P_{L} \cdot  A_{HB} &=& P_{L} \cdot Z \cdot  \frac {1}{1-K_{0}} \cdot  T_{0}
{}.
\label{AHB.equals.old}
\end{eqnarray}
The algebraic proof of this equation, starting from Eq.\
(\ref{operator.bare.H}), is left as an exercise. We can also define
the densities of heavy quarks by $P_H \cdot A_{HB} = P_H \cdot Z \cdot
\frac {1}{1-K_{0}} \cdot T_{0}$, but we will not need to use the
definition here, since only light parton densities appear in the
factorization theorem.

At first sight it appears that the bare parton densities $A_{HB}$ are
identical to those in the previous version of the factorization
theorem.  This is not quite so, because we are using a different
renormalization subscheme for the QCD action, both subschemes being
part of the CWZ \cite{CWZ} family of schemes.  Green functions in the
two subschemes differ by factors associated with the changes in the
wave function renormalization factors.  In addition, even without wave
function renormalization, the numerical values of the coefficients in
the perturbation expansion of $K_{0}$, etc., would differ because the
numerical value of the coupling $\alpha _{s}$ differs between the two
subschemes.  This can all be summarized by saying that $K_{0}$,
$T_{0}$ and $C_{0}$ in the two subschemes differ by a renormalization
group transformation.

When we renormalize the operators, and hence construct the
renormalized factorization theorem, we will need to work in terms of
$K_{0}$ rather than $K_{H}$.  So we rewrite our new coefficient
function $C_{HB}$ in terms of fully 2PI amplitudes.  This is done
quite simply by defining a new projection operator $Z_{H}$ that is
zero when applied to heavy quark lines and that is $Z$ on light parton
lines.  Then $Z_{H}=Z\cdot P_{L}$.

Graphically, the coefficient function $C_{HB}$ given in Eq.\
(\ref{coefficient.bare.H}) is $C_{0}$ with any number of $K_{0}$'s
attached.  If neighboring rungs are connected by active partons,
then a factor of $1-Z$ is inserted, but connections by heavy quarks
are left unaltered.  A straightforward but somewhat lengthy
algebraic derivation shows that Eq.\ (\ref{coefficient.bare.H})
implies that
\begin{equation}
    C_{HB} = C_{0} \cdot  \frac {1}{1 - (1-Z_{H})K_{0}}
          \cdot  Z \cdot P_{L} .
\label{coefficient.bare.H0}
\end{equation}
Observe that on an active light parton $1-Z_{H} = 1-Z$ and on a heavy
quark $1-Z_{H} = 1$, so that this equation agrees with the verbal
description given at the beginning of the paragraph.

\subsection{Renormalized factorization theorem}

Next we copy and slightly modify the steps needed to derive the
renormalized factorization theorem.  To define the renormalized
parton densities, we need to use a renormalization scheme in which
the heavy parton is treated as non-partonic.  So we define
\begin{eqnarray}
    A_{HR} &=& \sum _{n=0}^{\infty }
    Z \cdot  \left[ K_{0} \cdot  \left( 1 - \polepart_{H} \right)
        \right] ^{n}
    \cdot  T_{0}
\nonumber\\
    &=&
    Z \cdot  \frac {1}{1 - K_{0} \cdot  \left( 1 - \polepart_{H} \right)}
    \cdot  T_{0} .
\label{operator.ren.H}
\end{eqnarray}
The renormalization is defined by $\polepart_{H}$, which is an
operation that acts to the left. We define $L\polepart_{H}$ as
follows: If $L$ contains heavy quark loops and its rightmost external
lines are light partons, then $L\polepart_{H}$ is the value of
$L(q,k,M,m)$ when $k^{-}$ and ${\bf k}_{T}$ are replaced by zero and
the light parton masses $m$ are replaced by zero.  If $L$ contains no
heavy lines, then, $L\polepart_{H}$ is just the \MSbar{} pole part of
$L$.  The remaining case is when we apply $\polepart_{H}$ to graphs
with external heavy lines.  There is a choice of scheme that is not
determined by the overall requirements listed in Sec.\ \ref{sec:good}.
This is similar to the non-uniqueness found by Roberts and Thorne
\cite{RT2,RT1}.  We will choose to define the operation to be
pole-part subtraction, in the \MSbar{} style, as we did in a similar
situation when renormalizing the interactions.

In accordance with the dictates of the BPH approach to
renormalization, counterterms are kept with the graphs they subtract.
Thus $L\polepart_{H}$ is only used when $L$ is a quantity for which
all subdivergences have been subtracted.  This also ensures \cite{CWZ}
that the use of zero momentum subtractions for subgraphs containing
heavy quark loops introduces no IR divergences in the counterterms.

With these definitions, we can copy most of the previous derivation of
a renormalized factorization theorem.  First we observe the the
relation between renormalized and bare parton densities has the form
\begin{equation}
   A_{HR} = G_H \otimes A_{HB} ,
\end{equation}
where we use
\begin{equation}
    G_H \equiv  Z -
        Z \cdot  \frac {1}{1 - K_{0}
        \cdot  \left( 1 - \polepart_H \right)}
        \cdot  K_{0} \polepart
\end{equation}
instead of $G$ given by Eq.\ (\ref{ren.factor}).  Then we express the
factorization theorem in terms of renormalized parton densities
\begin{equation}
   F = C_{HR} \otimes A_{HR} + \mbox{remainder, $r_H$} ,
\label{factorization.ren.H}
\end{equation}
where the renormalized coefficient function is
\begin{equation}
   C_{HR} = C_{HB} \otimes G_H^{-1} .
\label{coefficient.ren.H}
\end{equation}

Finally, we bring in the decoupling theorem.  This implies that a
renormalized graph for $A_{HR}$ is suppressed by a power of
$\Lambda/M$ if it contains any heavy quark lines.  This is a
consequence of the use of a renormalization scheme which obeys
manifest decoupling, for both the interaction and operator matrix
elements.  We are assuming here that the target hadron in the
structure function is a light hadron.  One case of this result is that
the density of a heavy quark is power suppressed in the scheme we are
using in this section.  This result only applies to the renormalized
heavy quark density, not to the bare heavy quark density.

We can therefore restrict the renormalized coefficient function so
that its external lines are light, and the factorization theorem becomes
\begin{equation}
   F = C_{HR} \otimes A_{LR} 
   + \mbox{power-suppressed remainder} ,
\label{factorization.ren.H1}
\end{equation}
where now the parton densities $A_{LR}$ are renormalized parton
densities in the effective low energy theory with the heavy quark
omitted.  There appears to be no simple formula for the remainder, and
notice that the remainder is {\em not} equal to $r_H$ defined in Eq.\
(\ref{remainder.H}).

As before, there appears to be no simple formula for the coefficient
function, but a simple recursion formula does exist and it corresponds
to the algorithms actually used to do calculations.  The formula is
almost the same as the previous one, Eq.\
(\ref{coefficient.recursion}):
\begin{equation}
    C_{HR}^{(n)} = 
    F_{p}^{(n)} - \sum _{j=0}^{n-1} C_{HR}^{(j)} A_{LRp}^{(n-j)} .
\label{coefficient.recursion.H}
\end{equation}
The structure function is to be computed on a light-parton target
only, not on a heavy target, and the light-parton masses are to be set
to zero.  The parton density has subscripts $LRp$, whose meaning is as
follows: The $L$ indicates that $A_{LRp}$ is computed with the
omission of all graphs containing heavy-quark lines.  The $R$
indicates that it is renormalized, and the $p$ indicates the same
(zero-mass light-parton) target as for the structure function.

The one complication in proving Eq.\ (\ref{coefficient.recursion.H})
results from the fact that in deriving the factorization theorem, Eq.\
(\ref{factorization.ren.H1}) on a general target, we omitted graphs
for $A_{LR}$ that contain heavy lines, but without giving a formula
for the omitted terms.  So the recursion formula Eq.\
(\ref{coefficient.recursion.H}) could be in error by similar terms,
i.e., there might be a power-suppressed remainder term on the
right-hand side.  In fact all graphs for $A_{LRp}$ that include heavy
quark lines are exactly zero when combined with their counterterms.
This is because they are being evaluated with their external momenta
at exactly the subtraction point.  Hence Eq.\
(\ref{coefficient.recursion.H}) is exact.

\subsection{Differences between heavy and light factorization}

The renormalized factorization theorem with heavy quarks, Eq.\
(\ref{factorization.ren.H1}), differs from the first factorization
theorem Eq.\ (\ref{factorization.ren}) in two respects:
\begin{enumerate}

\item
    The sum over partons in the heavy quark factorization is
    restricted to light partons only.

\item
    The parton densities differ by a change of scheme.

\end{enumerate}
The first point accounts for our terminology of contrasting
``active'' (or ``partonic'') with ``non-partonic'' quarks.  In the
factorization we derived for $Q \gtrsim M$, Eq.\
(\ref{factorization.ren}), the heavy quark is partonic: there is
a term involving hard scattering off a heavy quark.  In contrast,
in the factorization for $Q \lesssim M$, Eq.\
(\ref{factorization.ren.H1}), there is no such term.

There is an overlapping domain of utility of the two schemes.  This is
where both $Q$ and the \MSbar{} scale $\mu $ are of order $M$.  In
this situation there are no large logarithms in the coefficient
functions and no large logarithms in the coefficients that relate the
two schemes.  This overlap is important because it implies that the
relation between the parton densities in the two schemes can be
computed perturbatively.  In practical applications it should be
remembered that at large $x$, the physical threshold for heavy-quark
production can be well above $Q$, and consequently the region where
the two schemes have common domains of utility should then be
appropriately biased upwards in $Q$.

When the heavy quark is treated as non-partonic, its parton density is
not used in the factorization theorem, and one might suppose that the
heavy quark density does not exist at all.  In fact the heavy quark
density does exist, because one can define it by exactly the usual
operator formula, together with renormalization (as dictated by the
CWZ scheme).  The important fact is that the heavy quark (and
antiquark) densities can be expressed in terms of the light parton
densities by a version of factorization.  This is a heavy quark
expansion for matrix elements of heavy-quark operators in light
states, and the argument was first given by Witten \cite{Witten} for
the case of local operators.  In the subscheme where the heavy quark
is non-partonic, the result is quite simple: the heavy quark densities
are suppressed by a power of the heavy quark mass:
\begin{equation}
   f_{H/p} = O(\Lambda /M) .
\label{H.from.L.true}
\end{equation}
We used this property in our derivation of the factorization theorem.

\section{Multiple heavy quarks}
\label{sec:multiple}

Let us now suppose that we have the most general case that there
are several heavy quarks, whose masses may or may not differ
greatly, and that $Q$ can vary over a wide range.

\subsection{Factorization}

In this situation, we define a series of subschemes, each of which is
labeled by the subset of the flavors of quarks and gluon which
are treated as active (or partonic). The other flavors in the
subscheme we call non-partonic.  The choice of subscheme is made
according to the value of $Q$.  If $Q$ is much larger than the
mass of a particular quark, then that quark is partonic.  If $Q$
is much smaller than the mass of a particular quark, then that
quark is non-partonic.  If $Q$ is comparable to the mass of a
particular quark, we may freely choose whether the quark is
partonic or non-partonic.  Gluons are light, so they are always
partonic.   We can define the scheme by saying that the $n_{A}$
lightest quarks are partonic.

Factorization is derived by a minor extension of the procedure in
Sec.\ \ref{sec:small.Q}.  In that section we had one heavy quark,
which was treated as non-partonic, with the gluon and other quarks
being treated as partonic.  We simply need to replace all references
to a ``heavy quark'' by references to ``non-partonic quarks''.  Thus
renormalization counterterms are generated by \MSbar{} pole terms,
except for mass renormalization of heavy quarks, which is always
performed on shell, and except for graphs with loops of non-partonic
quarks, whose counterterms are computed at zero external momentum and
with the masses of the active partons set to zero.  This defines the
appropriate version of the renormalization operator that is to replace
$\polepart$ in Eq.\ (\ref{operator.ren}) or $\polepart_{H}$ in Eq.\
(\ref{operator.ren.H}).

In the construction of the coefficient function and the remainder
for the factorization formula, the operation $Z_{H}$ must be
replaced by $Z_{n_{A}}$, which is $Z$ when applied on an active quark
or gluon and zero on non-partonic quarks. 

The methods used to construct the two factorization proofs
readily generalize to show that the remainder is suppressed by a
power of $Q$, provided that all the active partons have masses
less than or of the order of $Q$.  Moreover, in the perturbative
expansion of the coefficient function, if the \MSbar{} scale $\mu $
is of order $Q$, there will be no large logarithms of ratios of
$Q$, $\mu $ and quark masses provided also that the masses of the
non-partonic quarks are all larger or comparable with $Q$.  The
coefficient functions have infra-red-safe limits when masses of
active partons are set to zero.  (This applies in particular to
the light quarks; their masses may always be set to zero in the
coefficient functions.)

\subsection{When can the masses of active partons be set to zero?}

The setting to zero of active parton masses in the renormalization
prescription is necessary to get the simplest results, for example for
the renormalization-group coefficients.  It is always legitimate.

Moreover, if one is computing the coefficient function for a
particular external quark, then one can set to zero the mass of this
quark and of the lighter partons, as explained around Eq.\
(\ref{Z1.def.heavy}).  It is only with this prescription that the
recursion formula for the coefficient function, Eq.\
(\ref{coefficient.recursion.H}) is exact.

As an example, suppose that one is treating the charm quark (of mass
$m_c = 1.5\,{\rm GeV}$) as partonic but the bottom quark (of mass $m_b
= 4.5\,{\rm GeV}$) as non-partonic.  This implies that we are treating
phenomena on a scale of at least $m_c$. Furthermore, suppose that one
has decided that the charm quark is not sufficiently light compared to
$m_b$ for its mass to be neglected.  Then in coefficient functions
with external gluons, for example, one leaves both the masses of the
charm and bottom quarks at their physical values.  In contrast, in a
coefficient function with an external charm quark, its mass may be set
to zero.  As explained around Eq.\ (\ref{Z1.def.heavy}), this may be
done without loss of accuracy, since any errors are taken care of by
higher-order coefficients with lighter external partons.

\section{Matching conditions and evolution equations}
\label{sec:match.evolve}

\subsection{Matching conditions}
\label{sec:match}

As a consequence of the decoupling theorem, the density of a {\em
non}-partonic quark is suppressed by a power of $\Lambda /M$, where
$M$ is the mass of the quark, so we will normally approximate these
densities by zero.

Furthermore there are matching conditions between the parton densities
with $n_{A}$ and $n_{A}+1$ active quarks.  The coefficients relating
the parton densities are functions of the quark masses and $\mu $, and
have no large logarithms provided that $\mu $ is of the order of the
mass of quark $n_{A}+1$.  The coefficients also have infra-red-safe
limits when the masses of the $n_{A}$ lightest quarks are set to zero.

These matching conditions, have been given in Ref.\ \cite{ACOT} to
order $\alpha _{s}$ and in Ref.\ \cite{BMSvN} to order $\alpha
_{s}^{2}$.  They are applied to calculate the parton densities with
$n_{A}+1$ active quarks from the parton densities with $n_{A}$ active
quarks.  The conditions are to be applied at a value of the
renormalization scale around the mass of quark $n_{A}+1$.  Given that
we set the density of the quark $n_{A}+1$ to zero when it is
non-partonic, the matching conditions give initial values for all
$n_{A}+1$ quarks and the gluon which can therefore be evolved upward
in scale.  This gives an effective calculation of the density of quark
$n_{A}+1$ in the region where it is active.

\subsection{Evolution equations}
\label{sec:evolution}

We have a series of schemes labeled by the number of active
quarks, $n_{A} = 3, 4, 5, \dots$.  In each scheme we have densities
for the gluon and for each of the active quarks and antiquarks.
Up to power-suppressed corrections, the densities of the
non-partonic quarks and antiquarks are zero.  The active partons
evolve according to the standard DGLAP evolution equations, with
the kernels being those of the \MSbar{} with $n_{A}$ flavors.

\section{Miscellaneous comments}
\label{sec:comments}

\subsection{Relation to other methods of treating heavy quarks}

Calculations of heavy quark production often use what is called a
fixed-flavor-number scheme \cite{F2c,photo,DEN,had.heavy}.  This
corresponds exactly to the method described in the present paper if
the heavy quark is treated as non-partonic.  (For example, it
corresponds to a 3-flavor scheme for charm production and to a
4-flavor scheme for bottom production.)

Other calculations switch between different numbers of active quarks,
but neglect the masses of the active quarks in the coefficient
functions.  This is a valid approximation to the scheme here when
power corrections in $M/Q$ are negligible, but not when these power
corrections are important.  The scheme described in this paper does
not require the masses of active quarks to be neglected.

I have been unable to discover the justification of the scheme
proposed by Martin, Roberts, Ryskin and Stirling \cite{MRRS}.

Roberts and Thorne \cite{RT2,RT1} appear to have a scheme similar to
the one in the present paper.  But they do not present complete
proofs, and they make a number of incorrect or misleading
statements. For example, they state that ``the detailed construction
of the coefficient functions \dots is extremely difficult if not
impossible.''  As regards the general formalism, the construction is
exactly as difficult as in the light-quark case.  The only
computational complication is that in a calculation of the coefficient
functions, heavy quark masses must be retained.  All the necessary
Feynman-graph calculations for computing the coefficient functions at
order $\alpha _{s}^{2}$ have been done in Refs.\ \cite{BMSvN}, and all
that remains is to organize them to form the coefficient function by
use of the recursion relation Eq.\ (\ref{coefficient.recursion}).
This recursion relation is of the same form as the one used to obtain
the coefficient functions in the massless case.

\subsection{Modification of the schemes}
\label{sec:schemes}

It is possible to redefine the factorization results by a change of
scheme that defines the parton densities.  This is in effect a change
of the renormalization operation that defines them.

In addition, the details of the extraction of the asymptotics of the
structure functions may be changed by redefining the $Z$ operation.
The constraints on allowed redefinitions were explained in Sec.\
\ref{sec:redefinition.of.Z}, and they are implied by the requirements
for a good factorization scheme that were listed in Sec.\
\ref{sec:good}.  The $Z$ operations and the renormalization operation
should not change the validity of the error estimates used in the
proof of factorization.

I consider the \MSbar{} scheme to be the best underlying scheme
at the present state of the art, since it is the scheme most
commonly used for calculations of QCD corrections to hard
processes (at least when masses are ignored).

\subsection{Comparison with Zimmermann's approach}
\label{sec:Z}

One often gets the impression that Zimmermann's derivation
\cite{Zimmermann} of the operator product expansion (OPE) is
considered as the most reliable.  However, Zimmermann does not in fact
prove the results that we need for regular QCD phenomenology, even if
we restrict to the case that the OPE is sufficient.  (The derivation
in the present paper in fact applies to the Minkowski space structure
functions, rather than only to the integer moments of the
structure functions.  It is to these integer moments that the OPE in
its strict sense is restricted.)

His results suffer from two disadvantages.  The first is that his
Wilson coefficients have divergences in the zero-mass limit.  They are
not infra-red safe, and further work is needed to put the results in a
useful form for perturbative phenomenology in QCD.  The second
disadvantage is that his evolution equations are the inhomogeneous
Callan--Symanzik equations rather than the homogeneous
renormalization-group equations that can actually be used in practice.
The inhomogeneous term is not of a form susceptible to easy
calculation, so further work is needed to show that to a suitable
approximation, this term can be neglected.  In Tkachov's terminology
\cite{perfect}, Zimmermann's version of the OPE does not give a
``perfect asymptotic expansion'' at large $Q$. In contrast, the
factorization proved in the present paper is perfect in this sense.

In this section, we will see how Zimmermann's results can be
proved by our methods, and that they indeed suffer from the above
mentioned disadvantages.

The algebraic steps that led to our factorization theorem are shown in
Eq.\ (\ref{main.proof}).  The strategy in organizing the manipulations
was that the {\em right}-most factor of $Z$ should be made explicit.
Zimmermann's result can be obtained by arranging so that the {\em
left}-most $Z$ is picked out.  This results in the following
derivation:
\begin{eqnarray}
    F - r &=& C_{0} \cdot  \left[
                       \frac {1}{1-K_{0}} - (1-Z) \frac {1}{1 - K_{0}(1-Z)}
                   \right]
                 \cdot  T_{0}
\nonumber\\
          &=& C_{0} \cdot  \frac {1}{1-K_{0}} \cdot
                   \left[
                        1 - K_{0}(1-Z) - (1-K_{0})(1-Z)
                   \right]
                 \cdot  \frac {1}{1 - K_{0}(1-Z)}
                 \cdot  T_{0}
\nonumber\\
          &=& C_{0} \cdot  \frac {1}{1 - K_{0}} \cdot  Z
                 \cdot  \frac {1}{1-K_{0}(1-Z)} \cdot  T_{0} .
\label{Zimmermann.proof}
\end{eqnarray}
We therefore have a factorization theorem
\begin{equation}
   F  = C_{Z} \otimes A_{Z} + \mbox{non-leading power} ,
\label{factorization.Z}
\end{equation}
where the coefficient function is
\begin{equation}
   C_{Z} = C_{0} \frac {1}{1-K_{0}} Z ,
\label{coefficient.Z}
\end{equation}
and the operator matrix element is
\begin{equation}
   A_{Z} = Z \frac {1}{1-K_{0}(1-Z)} T_{0} .
\label{operator.Z}
\end{equation}
Notice first that the operator matrix element $A_{Z}$ in Zimmermann's
approach is already ultra-violet finite: the $1-Z$ factors in Eq.\
(\ref{operator.Z}) provide the necessary counterterms.  This is
contrast with our approach in Sect.\ \ref{sec:large.Q.proof}, where
some extra work was needed to express the factorization in terms of
renormalized operators. Unfortunately, the counterterms in
Zimmermann's approach are calculated at zero momentum, and so they
suffer from divergences in the massless limit, notably for the gluons.
Thus although the bare matrix elements (without renormalization) are
infra-red finite, if the hadron state is well behaved, the
renormalization procedure introduces mass divergences.

Moreover the coefficient function is ultra-violet finite, since it is
just a Green function of two currents and two partons.  In
Zimmermann's work, on the OPE, the external partons of the coefficient
function are given zero momentum; this corresponds to his use of zero
momentum subtractions to do renormalization.  The correct
generalization to Minkowski space problems is given by the operator
$Z$ defined in Eq.\ (\ref{Z.def1}): only the ``$-$'' and transverse
components of a momentum are set to zero.  Our derivation works
equally well with on-shell renormalization, with $Z$ defined by Eq.\
(\ref{Z.def.massive}).

However, Zimmermann's definition of the coefficient is not infra-red
finite.  One cannot set the masses to zero.  This is the strongest
reason for not regarding Zimmermann's approach as adequate for the
problems we are interested in.  It is a particular problem in QCD as
opposed to other field theories, since the gluon is intrinsically
massless.

\subsection{Other processes}
\label{sec:other}

Exactly the same methods that have been explained here can be
applied to other processes.  Also, if there turn out to be other
fields with color interactions, for example, squarks or gluinos, they
can be treated by minor generalizations of the same methods: we have
the choice of treating each massive field as either partonic or
non-partonic.

\section{Conclusions}
\label{sec:concl}

I have given a proof of factorization for deep-inelastic structure
functions including the effects of heavy quarks.  The methods are
general and include all non-leading logarithms.  The scheme
implemented is exactly that of ACOT \cite{ACOT}.  The proof is
applicable independently of the relative sizes of the heavy quark
masses and $Q$, and the size of the errors is a power of $\Lambda/Q$.
It can be readily extended to other hard processes.

Although this paper is quite lengthy, its core is really quite short.
The essential elements of the proof are:
\begin{enumerate}
\item
    Power counting is used to prove that the leading regions
    have the form symbolized by Fig.\ \ref{fig:leading.region}. This
    is a standard basic result of perturbative QCD.

\item
    The remainder, as defined in Eq.\ (\ref{remainder}), is then proved
    to be a non-leading power.  The proof is fairly obvious given the
    form of the leading regions.

\item
    The bare form of factorization then follows from the three lines
    of algebra given in Eq.\ (\ref{main.proof}).

\item
    Renormalization of the parton densities is implemented in
    Eq.\ (\ref{operator.ren}).  Then applying the inverse
    renormalization factor gives the renormalized factorization
    theorem.

\item
    Application of the factorization theorem to a parton target gives
    an algorithm for computing the coefficient function.

\end{enumerate}
This gives the factorization theorem when a heavy quark is treated as
partonic.  Simple modifications, plus the use of the decoupling
theorem, give the corresponding results when a heavy quark is
non-partonic.

When one is treating a heavy quark as partonic, it is valid to include
the heavy quark in the sum over partons in the factorization formula
even though it cannot really be treated as a parton, in Feynman's
sense.\footnote{ The word ``parton'' is used in two different senses
in this sentence!  } Errors in doing this are automatically taken care
of by the inclusion of higher-order terms in the coefficient
functions.  Since the heavy quark densities and the light parton
densities are of different sizes in the threshold region, a correct
leading-order calculation can only be done if lowest-order coefficient
functions are included for all possible subprocesses.  The
lowest-order coefficient functions are of different orders in
$\alpha_s$: The quark-induced processes have a lowest order $1$, and
the gluon induced process has a lowest order $\alpha_s$.  As $Q$
changes, the relative contributions of the different subprocesses
change in size. This mixing of orders is to be expected in any problem
where the parton densities have very different sizes, and is not
incorrect, contrary to the assertion of Roberts and Thorne \cite{RT1}.

Notice that there is an implicit unitarity sum over final states in
the whole of our work.  As explained on page \pageref{Unitarity.Sum},
this implies that the details of the final-state interactions do not
affect factorization or the calculation of the coefficient functions.
In particular, it is irrelevant that in Feynman-graph calculations,
there are on-shell partons in the final-state, even though in the
real-world there are only physical hadrons in the final-state.

\section*{Acknowledgments}

This work was supported in part by the U.S.\ Department of Energy
under grant number DE-FG02-90ER-40577.
I would like to thank CERN and DESY for their support during a sabbatical
leave where this work was initiated.
I would like to thank Fred Olness and Randy Scalise for some very
useful comments on a draft of this paper.


\appendix


\section*{Misleading derivation of formula for renormalized
         coefficient function}
\label{sec:bad.derivation}

In this appendix we show some apparently correct manipulations
can be used to justify a plausible but wrong formula for the
renormalized coefficient function.  The formula is
\begin{equation}
   C_{\rm cand} = C_{0} \cdot  \frac {1}{1- (1-Z) \cdot  K_{0}}
       \cdot  (1-\polepart ) \cdot  Z .
\label{coefficient.cand}
\end{equation}
(The subscript ``cand'' is to indicate this is a candidate for the
renormalized coefficient function.)  Expanded in powers of $K_{0}$
this gives
\begin{equation}
   C_{\rm cand} = C_{0} \cdot
       \sum _{n=0}^{\infty } [ (1-Z) K_{0} ]^{n}
       \cdot  (1-\polepart ) \cdot  Z .
\label{coefficient.cand.exp}
\end{equation}

This candidate coefficient function has some properties that make
it an obvious candidate for a renormalized coefficient function:
\begin{enumerate}

\item
    The factors of $1-Z$ prevent there from being leading
    contributions from regions where the momenta on the left are
    much higher in virtuality than those on the right.

\item
    This includes the case that the left-hand momenta are hard
    momenta, of virtuality of order $Q^{2}$, as in the leading
    regions Fig.\ \ref{fig:leading.region}, as well as the
    momenta that give ultra-violet divergences.

\item
    Thus the only leading regions are where all the momenta in
    $C_{\rm cand}$ are of virtuality of order $Q^{2}$ or where there is
    an ultra-violet divergence where all the momenta in some
    right-hand part of $C_{\rm cand}$ go to infinity.

\item
    The factor $1-\polepart$ cancels all the ultra-violet
    divergences.

\item
    The right-most factor of $Z$ defines the standard
    approximation appropriate to defining a hard-scattering
    coefficient that is coupled to a collinear target factor.

\end{enumerate}
Therefore $C_{\rm cand}$ represents an obvious way of applying
ultra-violet renormalization to the bare coefficient function
defined in Eq.\ (\ref{coefficient.bare}).

Let us now attempt to prove the factorization formula
\begin{equation}
    F = C_{\rm cand} \otimes A_{R} + \mbox{non-leading power}.
\end{equation}
The
following manipulations use just ordinary linear algebra,
together with the definitions of $C_{B}$, $A_{B}$, and $A_{R}$, and the
properties $Z^{2}=Z$ and $\polepart Z = \polepart$:
\begin{eqnarray}
    C_{\rm cand} \otimes A_{R}
    &=& C_{0} \frac {1}{1-(1-Z)K_{0}} (1-\polepart) Z G \otimes A_{B}
\nonumber\\
    &=& C_{B} (Z-\polepart)
        \left[
           Z - Z K_{0} \frac {1}{1-(1-\polepart)K_{0}} \polepart
        \right]
        A_{B}
\nonumber\\
    &=& C_{B}
        \left[
          Z-\polepart
          - (Z K_{0} - \polepart K_{0})
            \frac {1}{1-(1-\polepart)K_{0}} \polepart
        \right]
        A_{B}
\nonumber\\
    &=& C_{B}
        \left[
          Z-\polepart
          + (1 - K_{0} + \polepart K_{0} - 1 + K_{0} - Z K_{0})
            \frac {1}{1-(1-\polepart)K_{0}} \polepart
        \right]
        A_{B}
\nonumber\\
    &=& C_{B}
        \left[
          Z
          - (1 - (1-Z) K_{0})
            \frac {1}{1-(1-\polepart)K_{0}} \polepart
        \right]
        A_{B}
\nonumber\\
    &=& C_{B} \otimes A_{B}
        - C_{0}
            \frac {1}{1-(1-\polepart)K_{0}} \polepart
        A_{B}
\nonumber\\
    &=& C_{B} \otimes A_{B}
        - C_{0} \frac {1}{1-K_{0}} \left( 1-K_{0} \right)
            \frac {1}{1-(1-\polepart)K_{0}} \polepart
        A_{B}
\nonumber\\
    &=& C_{B} \otimes A_{B}
        - C_{0} \frac {1}{1-K_{0}} \left( 1 - (1-\polepart)K_{0}  -\polepart
K_{0}\right)
            \frac {1}{1-(1-\polepart)K_{0}} \polepart
        A_{B}
\nonumber\\
    &=& C_{B} \otimes A_{B}
        - C_{0} \frac {1}{1-K_{0}} \polepart
            \left[
               1
               - K_{0} \frac {1}{1-(1-\polepart)K_{0}} \polepart
            \right]
        A_{B} .
\label{false}
\end{eqnarray}
In the second term of the extreme right-hand side, we have a
pole-part operation applied to a quantity without ultra-violet
divergences, $C_{0}/(1-K_{0})$.  This second term is therefore zero,
and we appear to have proved $C_{\rm cand} \otimes A_{R} = C_{B} \otimes
A_{B}$,
which is sufficient to prove factorization, since $C_{B} \otimes A_{B}$
equals the structure function $F$, up to a power-suppressed
remainder.

Unfortunately, the above derivation is false.  It has assumed
that the operation of taking the pole part obeys all the rules of
linear algebra, including associativity.  The problem can be seen
at the first order in $K_{0}$.  There are two terms on the
left-hand side of Eq.\ (\ref{false}):
\begin{equation}
   \left[ C_{0} (1-Z) K_{0} (1-\polepart) Z \right]
   \left[ Z T_{0} \right]
+  \left[ C_{0} Z \right]
   \left[ Z K_{0} (1-\polepart) T_{0} \right] .
\label{false.1}
\end{equation}
The square brackets are used to delimit factors belonging to the
coefficient and to the operator.  The terms with a pole-part are
\begin{equation}
   - \left[ C_{0} (1-Z) K_{0} \polepart \right]
     \left[ Z T_{0} \right]
   - \left[ C_{0} Z \right]
     \left[ Z K_{0} \polepart T_{0} \right]
=
   \left[ C_{0} Z K_{0} \polepart \right] \left[ Z T_{0} \right]
   - \left[ C_{0} Z \right] \left[ Z K_{0} \polepart T_{0} \right] ,
\label{false.1.P}
\end{equation}
where we have observed (correctly) that $C_{0} K_{0}$ has no
ultra-violet divergence.

The two terms in Eq.\ (\ref{false.1.P}) appear to cancel.  In
fact this is not so.  We are taking
\begin{equation}
   \left[ \mbox{pole part}(C_{0} Z K_{0} ) \right ] T_{0}
   - C_{0} Z \left[ \mbox{pole part}(Z K_{0} ) \right ] T_{0} .
\label{false.1.diff}
\end{equation}
This is not, in general, zero, as can be seen by taking a simple
mathematical example.  Let us replace $C_{0}Z$ and $ZK_{0}$ by
\begin{eqnarray}
    C_{0}Z = 1 + a \epsilon , ~~~ ZK_{0} = \frac {1}{\epsilon } + b .
\end{eqnarray}
Then Eq.\ (\ref{false.1.diff}) becomes
\begin{eqnarray}
   \mbox{pole part}\left[
                      (1+a\epsilon )
                      \left(\frac {1}{\epsilon }+ b \right)
                    \right]
   - (1+a\epsilon ) \mbox{pole part}\left( \frac {1}{\epsilon } + b \right)
 &=& \frac {1}{\epsilon } - (1+a\epsilon ) \frac {1}{\epsilon }
\nonumber\\
 &=& -a,
\end{eqnarray}
which is clearly non-zero.

Treating $\polepart$ as an associative operator has failed.


\end{document}